\documentclass[aps,prb,twocolumn,superscriptaddress,
amsmath,amssymb]{revtex4-2}

\usepackage{graphicx}
\usepackage{dcolumn}
\usepackage{bm}
\usepackage{multirow}
\usepackage[colorlinks=true,linkcolor=blue,urlcolor=blue, citecolor=blue]{hyperref}
\usepackage{hyperref}

\graphicspath{{./Figures/}}

\begin{document}

\renewcommand{\arraystretch}{1.2}

\title{First-principles calculations of defects in metal halide perovskites: \\ a performance comparison of density functionals}

\author{Haibo Xue}
\affiliation{Materials Simulation and Modelling, Department of Applied Physics, Eindhoven University of Technology, P.O. Box 513, 5600MB Eindhoven, the Netherlands.}
\affiliation{Center for Computational Energy Research, Department of Applied Physics, Eindhoven University of Technology, P.O. Box 513, 5600MB Eindhoven, the Netherlands.}

\author{Geert Brocks}
\affiliation{Materials Simulation and Modelling, Department of Applied Physics, Eindhoven University of Technology, P.O. Box 513, 5600MB Eindhoven, the Netherlands.}
\affiliation{Center for Computational Energy Research, Department of Applied Physics, Eindhoven University of Technology, P.O. Box 513, 5600MB Eindhoven, the Netherlands.}
\affiliation{Computational Materials Science, Faculty of Science and Technology and MESA+ Institute for Nanotechnology, University of Twente, P.O. Box 217, 7500AE Enschede, the Netherlands.}

\author{Shuxia Tao}
\email{s.x.tao@tue.nl}
\affiliation{Materials Simulation and Modelling, Department of Applied Physics, Eindhoven University of Technology, P.O. Box 513, 5600MB Eindhoven, the Netherlands.}
\affiliation{Center for Computational Energy Research, Department of Applied Physics, Eindhoven University of Technology, P.O. Box 513, 5600MB Eindhoven, the Netherlands.}

\begin{abstract}
Metal halide perovskite semiconductors have outstanding optoelectronic properties. Although these perovskites are defect-tolerant electronically, defects hamper their long-term stability and cause degradation. Density functional theory (DFT) calculations are an important tool to unravel the microscopic structures of defects, but results suffer from the different approximations used in the DFT functionals. In the case of metal halide perovskites, qualitatively different results have been reported with different functionals, either predicting vacancy or interstitial point defects to be most dominant. Here, we conduct a comprehensive comparison of a wide range of functionals for calculating the equilibrium defect formation energies and concentrations of point defects in the archetype metal halide perovskite, MAPbI$_3$. We find that it is essential to include long-range Van der Waals interactions in the functional, and that it is vital to self-consistently optimize structure and volume of all compounds involved in the defect formation. For calculating equilibrium formation energies of point defects in MAPbI$_3$ and similar metal halide perovskites, we argue that the exact values of the chemical potentials of the species involved, or of the intrinsic Fermi level, are not important. In contrast to the simple Schottky or Frenkel pictures, we find that the dominant defects are MA and I interstitials, and Pb vacancies. 
\end{abstract}

\maketitle

\section{Introduction}\label{section:Introduction}
Metal halide perovskites are emerging photovoltaic materials for solar cells, whose power conversion efficiency (PCE) has increased spectacularly from a few to over 25 percent over the last decade \cite{Kojima2009, NREL2021}. The impressive photovoltaic performance is linked to the materials' defect tolerance \cite{Zakutayev2014, Steirer2016}, and attributed to a low intrinsic defect density, $10^{11}$-$10^{16}$ cm$^{-3}$, \cite{Adinolfi2016, Leijtens2016, Chen2016a, DeQuilettes2015, Blancon2016, Samiee2014} and to the dominant defects not creating deep trap levels \cite{Yin2014, Shi2015}. Despite the impressive progress, the long-term stability of the perovskite solar cells is still limited, which is ascribed to intrinsic instabilities of the material, and to defect-triggered degradation reactions \cite{Lindblad2014, Chen2014, Sadoughi2015, Steirer2016, Wang2017, Jiang2018, Fang2018, Motti2019, Li2019, Li2020}.

The defects in halide perovskites are rich in types and charge states. Taking the prime metal halide compound MAPbI$_3$ as an example, there are already six different simple point defects, i.e. vacancies and interstitials, and this number is doubled or tripled if one takes different charge states into consideration \cite{Yin2014}. To increase the long-term efficiency and stability of perovskite solar cells, it is necessary to come up with efficient strategies for defect passivation, which requires an understanding of the fundamental defect chemistry and physics of perovskites. It is therefore essential to identify the types and densities of defects, and their consequences for the electronic properties. 

In experiment, the overall density and distribution of trap states can be extracted through electrical measurements \cite{Samiee2014, Leijtens2016, Adinolfi2016}. However, it remains challenging to pin-point the exact types of defects and correlate those to electronic trap states \cite{Chen2019}. Another challenge emerging from experiments is that the types of dominant defects can depend on details of the fabrication process, such as the ratio of precursor materials used \cite{Yin2014, Son2016, Bi2016}, or the thermal annealing procedure \cite{Chen2014, Bi2014, Dualeh2014}, leading to different morphology and final compositions. This stresses the importance of being able to identify which defects would be present under ideal thermodynamic conditions.

To provide atomistic insight in the defect physics, computational modelling based upon density functional theory (DFT) is often invoked. In particular, one focuses on calculating the defect formation energies (DFEs) and charge state transition levels (CSTLs), which are key quantities for predicting the abundance and the electronic nature of defects \cite{Walle2004}. Unfortunately, the quality of density functionals is not yet at a stage that all key defect quantities can be obtained simultaneously with the same accuracy \cite{Freysoldt2014}. Besides total energy differences between pristine and defective structures, obtaining DFEs requires accurate calculations of the thermochemistry (formation enthalpies) of sometimes widely different compounds (metals, insulators, molecules), as well as of basic electronic properties, such as band edge positions and band gaps. 

A general-purpose semi-local functional based upon the generalized gradient approximation (GGA), such as the Perdew-Burke-Ernzerhof (PBE) functional \cite{Perdew1996}, can lead to errors in formation enthalpies of order $\sim 0.2$ eV/atom \cite{Tran2016, Zhang2018}. Hybrid functionals do not give a uniform improvement \cite{Tran2016, Zhangguoxu2018}, and in some cases, (transition) metals in particular, yield significantly worse formation enthalpies \cite{Stroppa2008}. Given the fact that using a hybrid functional increases the computational costs by one to two orders of magnitude, and the results depend on how and how much Hartree-Fock exchange is mixed in, hybrid functionals seem a less attractive option for calculating formation enthalpies at present. 

To deal with the imperfect formation energies resulting from PBE, semi-empirical schemes have been developed, such as the fitted elemental-phase reference energies (FERE) approach \cite{Stevanovic2012, Peng2013}. It assumes the error predominantly depends on the overall composition, and attempts an improvement through using the total energies of elemental phases as fitting parameters. The FERE approach can improve DFEs significantly \cite{Peng2013}, but it requires the construction of an extensive parameter set for each group of compounds studied.

A significant step forward is presented by the recently developed strongly constrained and appropriately normed (SCAN) meta-generalized gradient approximation (meta-GGA) functional \cite{Sun2015, Sun2016}. Compared to standard GGA, SCAN more uniformly improves formation enthalpies of both strongly bonded solids and molecules \cite{Tran2016, Zhangguoxu2018, Yang2019}, typically at least halving the error as compared to PBE \cite{Zhang2018}. 

A complicating factor in solids such as the hybrid metal-halide perovskites, is that heavy elements like Pb and I are involved, and organic cations are embedded in the inorganic metal-halide framework, both of which suggest that Van der Waals (vdW) interactions may play a non-negligible role in the bonding. The simplest way to include vdW interactions, is to add an atom-pairwise parametrized vdW energy to a standard DFT total-energy expression, such as in the D3 or D3(BJ) schemes \cite{Grimme2010,Grimme2011}, leading to the PBE-D3 and PBE-D3(BJ) functionals, for instance. Mixed results have been reported, where PBE-D3/D3(BJ) gives good results on the cohesive energy of rare-gas solids, for instance \cite{Tran2016}, but PBE-D3 does not perform well in cases where anion-anion interactions play a significant role \cite{Yang2019}.

A fundamentally different approach constructs an explicitly non-local density functional, which incorporates vdW interactions directly, where one has to carefully balance these non-local with (semi-)local terms in the functional. The rev-vdW-DF2\cite{Hamada2014, Klime2011, Dion2004, Guillermo2009} functional shows good results for weakly and strongly bound solids, with errors in formation energies on the scale of $\sim 0.1$ eV/atom, as does the SCAN+rVV10 functional, where the latter has the additional advantage that it is more accurate for molecules as well \cite{Tran2019}. 

The functionals mentioned above do not solve the band gap problem of DFT \cite{Freysoldt2014, Peng2013}. This problem is actually less severe if calculating DFEs under thermodynamic equilibrium conditions, as we will discuss below, which is the focus of the present paper. To calculate electronic properties such as CSTLs accurately on an absolute energy scale, requires the correct positions of the band edges. Post-processing DFT results using GW or hybrid functional calculations can in principle be applied for this purpose \cite{Peng2013, Meggiolaro2018review}, but in the present paper we will only compare pure DFT calculations.

Defect calculations on hybrid metal iodide perovskites using the PBE functional, have found the iodine vacancy to be one of the dominant point defects \cite{Michael2014, Buin2014, Buin2015, Ming2016}. In contrast, more recent studies incorporating vdW corrections using the DFT-D3 scheme, as well as a hybrid functional post-processing step, have led to the conclusion that the iodine interstitial is a more dominant point defect \cite{Meggiolaro2018-iodine-chemistry, Meggiolaro2018review, Meggiolaro2018_I2}. These studies illustrate the sensitivity of the predicted defect properties to the functionals used, and call for a careful and systematic comparison. 

Whereas the addition of semi-empirical atom-pairwise vdW corrections within the D3 approach \cite{Grimme2010} have a sizable effect on the calculated DFEs of MAPbI$_3$ \cite{Meggiolaro2018review}, an improved version based on the same scheme with Becke and Johnson damping added (DFT-D3(BJ)) \cite{Grimme2011}, which provides better corrections for non-bonding distances and intramolecular interactions \cite{Pols2021}, has not been tried yet. The same holds for accurate non-local vdW density functionals, such as rev-vdW-DF2. 

The SCAN functional is reported to perform better than standard GGA functionals in predicting the orientation of organic cations in halide perovskites and the phase transitions of these materials \cite{Bokdam2017, Lahnsteiner2018}, but has not yet been applied in calculating defects of halide perovskites. The non-local correlation functional rVV10 seamlessly supplements SCAN with long-range vdW interactions \cite{Sabatini2013}, and the resulting SCAN+rVV10 functional \cite{Peng2016} can also be tested in defect calculations on hybrid metal halide perovskites. Finally, as these perovskites contain heavy elements like Pb and I, the effects of spin-orbit coupling (SOC) should also be considered \cite{Even2013}.

In the present work, we use a wide range of functionals to calculate the DFEs and CSTLs of the intrinsic vacancies and interstitials in MAPbI$_3$, including PBE, PBE-SOC, SCAN, PBE-D3, PBE-D3(BJ), rev-vdW-DF2 and SCAN+rVV10. Comparing all data sets to the ones obtained by using PBE, the effects of switching from GGA to meta-GGA, from pairwise atomic to non-local vdW corrections, with and without SOC, are analyzed. While SOC has only little effect on the structures and the energies, it does impact the positions of the CSTLs. However, the latter can be straightforwardly included by applying an post-correction. 

In contrast, the inclusion of vdW interactions is shown to have a large effect on the DFEs, where the PBE-D3(BJ), rev-vdW-DF2 and SCAN+rVV10 functionals give similar results. To obtain accurate results, we find that it is vital to optimize structures self-consistently for each functional. The SCAN functional captures part  of the non-chemical bonding interactions, but including long-range vdW interactions, as in SCAN+rVV10, still has a sizable effect on the DFEs. Considering that it is a universally applicable functional for accurate calculations on solids and molecules, we suggest to use SCAN+rVV10 also for studying defects in other perovskite compounds.

\section{Computational Approach}\label{section:Computational_Methods}

\subsection{DFT calculations}
DFT calculations are performed with the Vienna ab initio simulation package (VASP) \cite{Kresse1993, Kresse1996, Kresse1996a}. Defects are created starting from a 2$\times$2$\times$2 tetragonal supercell of the MAPbI$_3$ perovskite structure with optimized volume and ionic positions. For each type of defect, cation or anion interstitial or vacancy, several different defective structures are tested, inserting or removing a cation or an anion in a different nonequivalent position, and optimizing the ionic positions within the supercell. The structures with the lowest total energy are then selected, and will be discussed below. A more detailed description of the strategy for creating the defective structures is given in the Supplemental Material, Figures S2-S4. 

These calculations are repeated for the following functionals: PBE \cite{Perdew1996}, SCAN \cite{Sun2015}, PBE-D3 \cite{Grimme2010}, PBE-D3(BJ) \cite{Grimme2011}, rev-vdW-DF2 \cite{Hamada2014, Klime2011, Dion2004, Guillermo2009} and SCAN+rVV10 \cite{Peng2016}, using the default parameters appropriate for the respective functionals. To assess the influence of spin-orbit coupling, we have also tested PBE+SOC \cite{Even2013}, as well as SCAN+SOC, but this time only on PBE- and SCAN-optimized structures, respectively. Our calculations use a plane wave kinetic energy cutoff of 500 eV and a $\Gamma$-point only $\mathbf{k}$-point mesh. The energy and force convergence criteria are set to 10$^{-4}$ eV and 0.02 eV/\AA, respectively.  

\subsection{Defect formation energies}
The defect formation energy (DFE) $\Delta H_{f}$ is calculated from
\begin{equation}\label{eq:DFE}
    \begin{aligned}
        \Delta H_{f}(D^q) = &E_\mathrm{tot}(D^q)-E_\mathrm{bulk} - \sum_{i} n_{i} \mu_{i} \\ 
        &+ q(E_{F}+ E_\mathrm{VBM}+ \Delta V) + E^q_\mathrm{corr}.
    \end{aligned}
\end{equation}
Here $D$ indicates the type of defect, interstitial or vacancy, and $q$ is its charge. $E_\mathrm{tot}(D^q)$ and $E_\mathrm{bulk}$ are the DFT total energies of the defective supercell and the pristine supercell, respectively. As it only makes sense to compare total energies of systems with the same atomic and electronic contents, and this is clearly not the case comparing a defective with a pristine cell, one has to define reservoirs to make up for the surplus or shortage of atoms and electrons of the defective cell. Here, $n_i$ and $\mu_i$ define the number of atoms and the chemical potential of the species $i$ added to ($n_i>0$) or removed from ($n_i<0$) the pristine supercell in order to create the defective supercell.

Likewise, the Fermi level $E_F$ defines the electrochemical potential for electrons, if electrons need to be added or removed from the pristine supercell to create the charge $q$. Ordinarily one chooses the zero of $E_F$ at the valence band maximum (VBM), so its energy $E_\mathrm{VBM}$ appears in the expression of Eq. (\ref{eq:DFE}). The VBM is typically hard to identify in a calculation of a defective cell, so one uses the value obtained from the pristine cell, shifted by $\Delta V$, which is, for instance, calculated by lining up the core level on one same atom in the pristine and the defective cell that is far removed from the defect. 

Finally, $E^q_\mathrm{corr}$ is introduced to correct for the electrostatic interaction between a charged point defect and its periodically repeated images. In agreement with Ref. \onlinecite{Meggiolaro2018review} we find that the 2$\times$2$\times$2 tetragonal supercell and the dielectric screening in MAPbI$_3$ are in fact sufficiently large, such that this correction is small and can be neglected.

The chemical potentials $\mu_i$ of the atomic species depend upon the growth and environmental conditions, but they are subject to some constraints. The conditions must be such that the MAPbI$_3$ phase is stable. In addition, we suppose it is in equilibrium with the PbI$_2$ phase. Moreover, the MAI phase is not supposed to form. This leads to the following relations
\begin{subequations}\label{eq:Chemical_Stability}
    \begin{align}
        &\mu _\mathrm{MA} + \mu _\mathrm{Pb} + 3 \mu _\mathrm{I} = \mu_\mathrm{MAPbI_3}, \label{eq:stability_of_MAPbI3}\\  
        &\mu_\mathrm{Pb} + 2 \mu_\mathrm{I} = \mu_\mathrm{PbI_2}, \label{eq:boundary_of_PbI2}\\
        &\mu_\mathrm{MA}  + \mu_\mathrm{I} < \mu _\mathrm{MAI}. \label{eq:boundary_of_MAI}
    \end{align}
\end{subequations}

All chemical potentials used in this work are represented by DFT total energies of the compounds involved, see Supplemental Information Table S4. The combination of Eqs. (\ref{eq:stability_of_MAPbI3}) and (\ref{eq:boundary_of_PbI2}) leaves one free parameter. Choosing $\mu _\mathrm{I}=\frac{1}{2}\mu_\mathrm{I_2,molecule}$, where $\mu_\mathrm{I_2,molecule}$ is the energy of a I$_2$ molecule, defines I-rich conditions. We have chosen this reference state to conform to previous calculations \cite{Meggiolaro2018-iodine-chemistry, Meggiolaro2018review, Meggiolaro2018_I2}. One may argue that the energy of a I$_2$ molecule in a bulk iodine environment, $\mu_\mathrm{I_2,bulk}$, is a more appropriate reference state. However, as we will show below, the exact choice of the iodine reference state is not so important when studying defects under intrinsic conditions. Choosing $\mu _\mathrm{I}=\frac{1}{2}\left(\mu_\mathrm{PbI_2} - \mu_\mathrm{Pb,bulk} \right)$, where $\mu_\mathrm{Pb,bulk}$ is the energy of a Pb atom in bulk Pb metal, defines I-poor conditions. 

The presence of molecule I$_2$ (I-rich) or bulk Pb (I-poor) represent rather extreme conditions, neither of which are representative for the growth and environmental conditions of MAPbI$_3$. We focus on so-called I-medium conditions, which are halfway between I-rich and I-poor, and thus defined by
\begin{equation}
\mu _\mathrm{I} = \frac{1}{4} \left( \mu_\mathrm{I_2,molecule} + \mu_\mathrm{PbI_2} - \mu_\mathrm{Pb,bulk} \right).
\label{eq:mu_iodine}
\end{equation}
The chemical potentials of Pb and MA can then be obtained from Eqs. (\ref{eq:boundary_of_PbI2}) and (\ref{eq:stability_of_MAPbI3}), respectively, whereas the constraint of Eq. (\ref{eq:boundary_of_MAI}) is then automatically obeyed.

Given the DFEs, the concentration of each type of defect $c(D^q)$ can be estimated by Boltzmann statistics
\begin{equation} \label{eq:defect_concentration}
    c(D^q)=c_0(D^q)  \exp(- \frac{\Delta H_f (D^q)}{k_BT}),
\end{equation}
where $c_0(D^q)$ is the density of possible sites for the defect (see Supplemental Information Table S3), $T$ is the temperature, and $k_B$ is the Boltzmann constant. The concentration $c(D^q)$ of charged defects ($q\neq0$) is obviously a function of the Fermi level $E_F$, see Eqs. (\ref{eq:DFE}) and (\ref{eq:defect_concentration}). 

\subsection{Intrinsic conditions}
If no charges are injected in a material, it has to be charge neutral, as expressed by
\begin{equation} \label{eq:charge_neutrality}
    p - n + \sum_{D^q} q c(D^q) = 0,
\end{equation}
where $p$ and $n$ are the intrinsic charge carrier densities of electrons and holes of the semiconductor material, and the sum is over all types of charged defects. Here $p$, $n$, and $c(D^q)$ are functions of $E_F$, so the charge neutrality condition, Eq. (\ref{eq:charge_neutrality}), serves to determine the intrinsic position of the Fermi level $E_F^{(i)}$. Note that, if $E_F^{(i)}$ is sufficiently far from the band edges, then $p$ and $n$ are small, and can be neglected.

Under intrinsic conditions, defined by Eq. (\ref{eq:charge_neutrality}), the DFEs and concentrations of the most prominent defects turn actually to be independent of the iodine chemical potential $\mu_\mathrm{I}$ over a large range of the latter \cite{Yang2015}. If one changes $\mu_\mathrm{I}$ to $\mu_\mathrm{I}+\delta$, then at a fixed Fermi level $E_F$ the DFE of an iodine interstitial changes from $\Delta H_{f}(\mathrm{I}_i)$ to $\Delta H_{f}(\mathrm{I}_i)-\delta$, Eq. (\ref{eq:DFE}). Because of the equilibria of Eqs. (\ref{eq:boundary_of_PbI2}) and (\ref{eq:stability_of_MAPbI3}), the DFEs of Pb and MA interstitials then change to $\Delta H_{f}(\mathrm{Pb}_i)+2\delta$ and $\Delta H_{f}(\mathrm{MA}_i)+\delta$, respectively. Likewise, under the same conditions the DFEs of I, Pb, and MA vacancies charge to $\Delta H_{f}(\mathrm{V_I})+\delta$, $\Delta H_{f}(\mathrm{V_{Pb}})-2\delta$, and $\Delta H_{f}(\mathrm{V_{MA}})-\delta$, respectively.

Close to intrinsic conditions, point defects in MAPbI$_3$ turn out to have a preference for a specific charge state, i.e., $\mathrm{I_i}^-$, $\mathrm{Pb_i}^{2+}$, $\mathrm{MA_i}^+$, $\mathrm{V_I}^+$, $\mathrm{V_{Pb}}^{2-}$, $\mathrm{V_{MA}}^-$. Following the discussion of the previous paragraph, if one changes the iodine chemical potential from $\mu_\mathrm{I}$ to $\mu_\mathrm{I}+\delta$, the DFEs at a fixed Fermi level $E_F$ of these point defects change to $\Delta H_{f}(D^q) + q\delta$, where $q$ is the charge of the defect. If one now applies the charge neutrality condition, Eq. (\ref{eq:charge_neutrality}), assuming that the Fermi level stays sufficiently far from the band edges such that $p\approx n \approx 0$, then the intrinsic Fermi level changes to $E_F'^{(i)} = E_F^{(i)}- \delta$. 

This means that, according to Eq. (\ref{eq:DFE}), the change in intrinsic Fermi level compensates exactly for the change in iodine chemical potential for the DFEs of the mentioned point defects, and the DFEs remain unaltered. In other words, the DFEs and concentrations of these defects are independent of the iodine chemical potential. 

\begin{figure}
    \includegraphics[width=1\columnwidth]{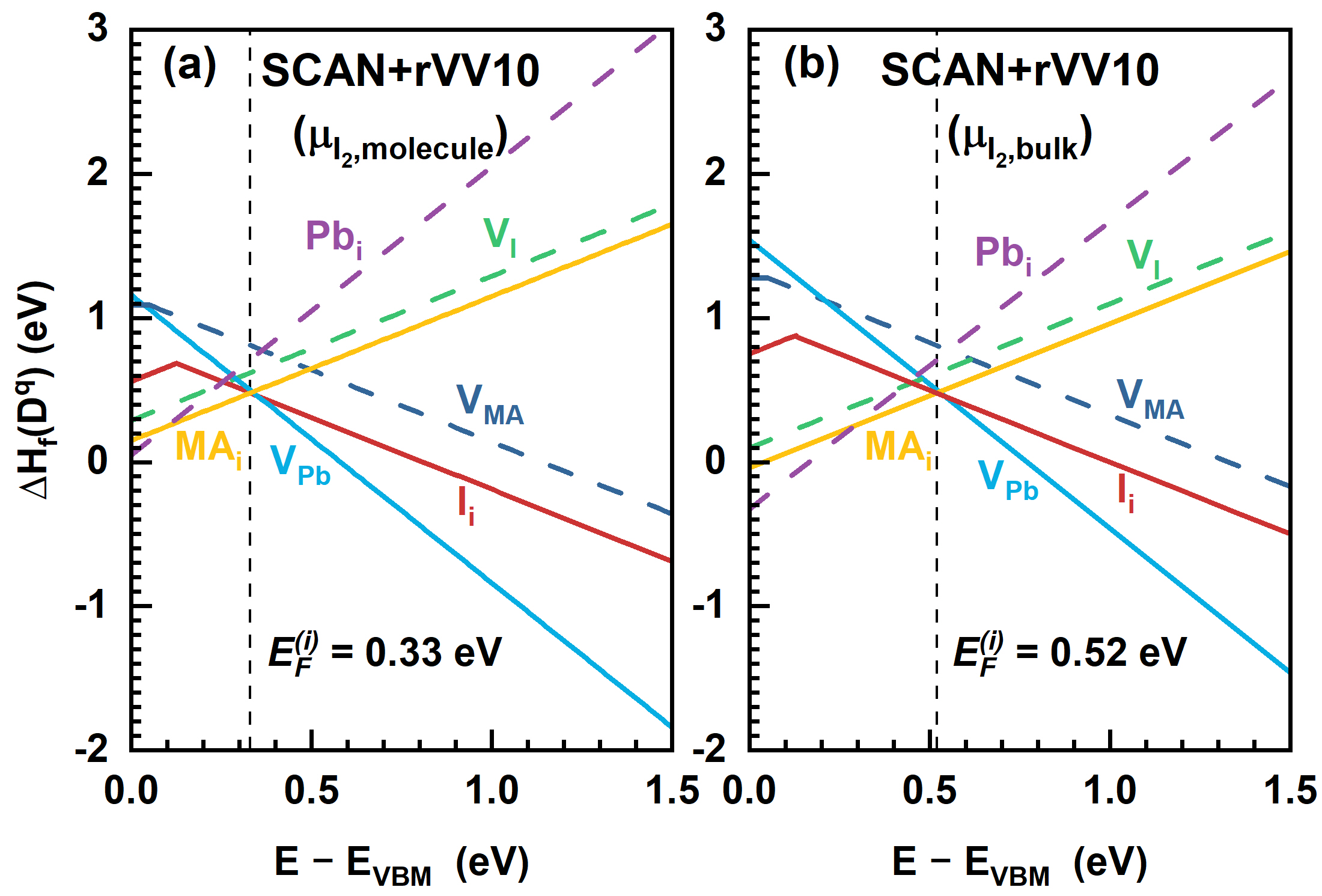}
    \caption{Defect formation energies, Eq. (\ref{eq:DFE}), calculated using (a) $\mu_\mathrm{I_2,molecule}$ in Eq. (\ref{eq:mu_iodine}), or (b) $\mu_\mathrm{I_2,bulk}=\mu_\mathrm{I_2,molecule} + 4\delta$, where  $\delta = -0.19$ eV. The vertical dotted line indicates the intrinsic Fermi level $E_F^{(i)}$, calculated from Eq. (\ref{eq:charge_neutrality}). Note that the difference in $E_F^{(i)}$ between (a) and (b) is $-\delta$, and the DFEs at $E_F^{(i)}$ are unchanged. The DFEs are calculated using the SCAN+rVV10 functional. \label{figure:DFE_chempot}}
\end{figure}

This effect is illustrated in Fig. \ref{figure:DFE_chempot}, showing the DFEs calculated with the SCAN+rVV10 functional, using $\mu_\mathrm{I_2,molecule}$ or $\mu_\mathrm{I_2,bulk}$, respectively in Eq. (\ref{eq:mu_iodine}). Note that this does not mean that the DFEs of all charge states are unaltered. For instance, $\Delta H_{f}(\mathrm{I}_i^+)$ becomes $\Delta H_{f}(\mathrm{I}_i^+)-2\delta$ (compare the upward sloping red lines in Fig. \ref{figure:DFE_chempot}(a) and (b)). For MAPbI$_3$ under normal equilibrium conditions, these particular charged states do not play a role, however.

\begin{figure}
    \includegraphics[width=1\columnwidth]{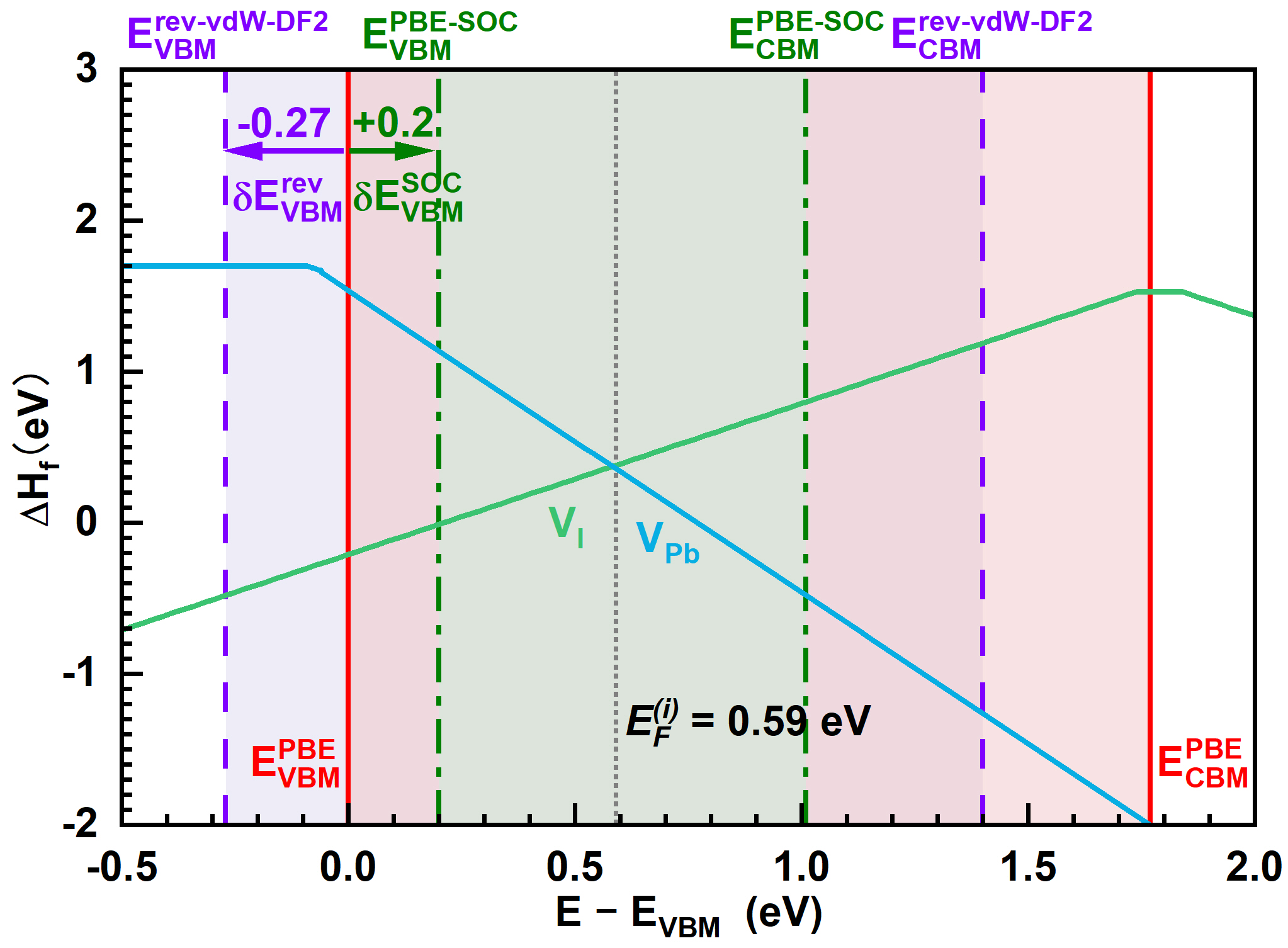}
    \caption{Defect formation energies of the two most prominent defects calculated with the PBE functional, as a function of $E_F$. The band edges (VBM and CBM) are indicated by red solid vertical lines. Adding spin-orbit coupling (SOC) decreases the band gap, and shifts the band edges  (green dashed-dotted lines), as does replacing the functional by rev-vdW-DF2 (purple dashed lines). Although the intrinsic Fermi level $E_F^{(i)}$ relative to the VBM (black dotted line), is different in all three cases, the (PBE-calculated) DFEs at $E_F^{(i)}$ remain the same.\label{figure:DFE_bandedges}}
\end{figure}

At intrinsic conditions, Eq. (\ref{eq:charge_neutrality}), the DFEs are also independent of the size of the band gap, or indeed of the exact positions of the band edges (VBM and CBM). To illustrate this, Fig. \ref{figure:DFE_bandedges} shows the band edge positions calculated with the PBE-SOC and rev-vdW-DF2 functionals, relative to those calculated with the PBE functional, lining up the 1s core level of a carbon atom on a MA molecule. Because of the difference in VBM, the intrinsic Fermi level $E_F^{(i)}$ relative to the VBM, is different in all three cases. However, the DFEs at $E_F^{(i)}$ (calculated with the respective functional) are the same for all three functionals. This is the basis of a posteriori band edge corrections for defect calculations \cite{Peng2013}. Of course, using a different functional for calculating the DFEs, the latter can still change, which will be discussed in Sec. \ref{section: results and discussion}. 

\subsection{Charge state transition levels}
Under non-equilibrium, operating, conditions, charges are injected in the material, shifting the position of the (quasi) Fermi level(s). The charge state transition level (CSTL) $\varepsilon(q/q')$ is defined as the Fermi level position where the charge states $q$ and $q'$ of the same type of defect have equal formation energy, $\Delta H_{f}(D^q)=\Delta H_{f}(D^{q'})$, so that if the Fermi level crosses the CSTL, the defect changes its charge state. As the DFEs have a simple linear dependence on $E_F$, Eq. (\ref{eq:DFE}), this condition translates into
\begin{equation} \label{eq:CSTL}
    \varepsilon(q/q')= \frac{\Delta H_{f}(D^q,E_{F}=0)-\Delta H_{f}(D^{q'},E_{F}=0)}{q'-q},
\end{equation}
where $\Delta H_{f}(D^q,E_{F}=0)$ is the defect formation energy calculated for $E_F = 0$. In calculating CSTLs, the absolute position of the VBM obviously plays a role.

\section{Results and discussion}\label{section: results and discussion}

\subsection{PBE calculations}\label{section:PBEresults}

\begin{figure}
    \includegraphics[width=1\columnwidth]{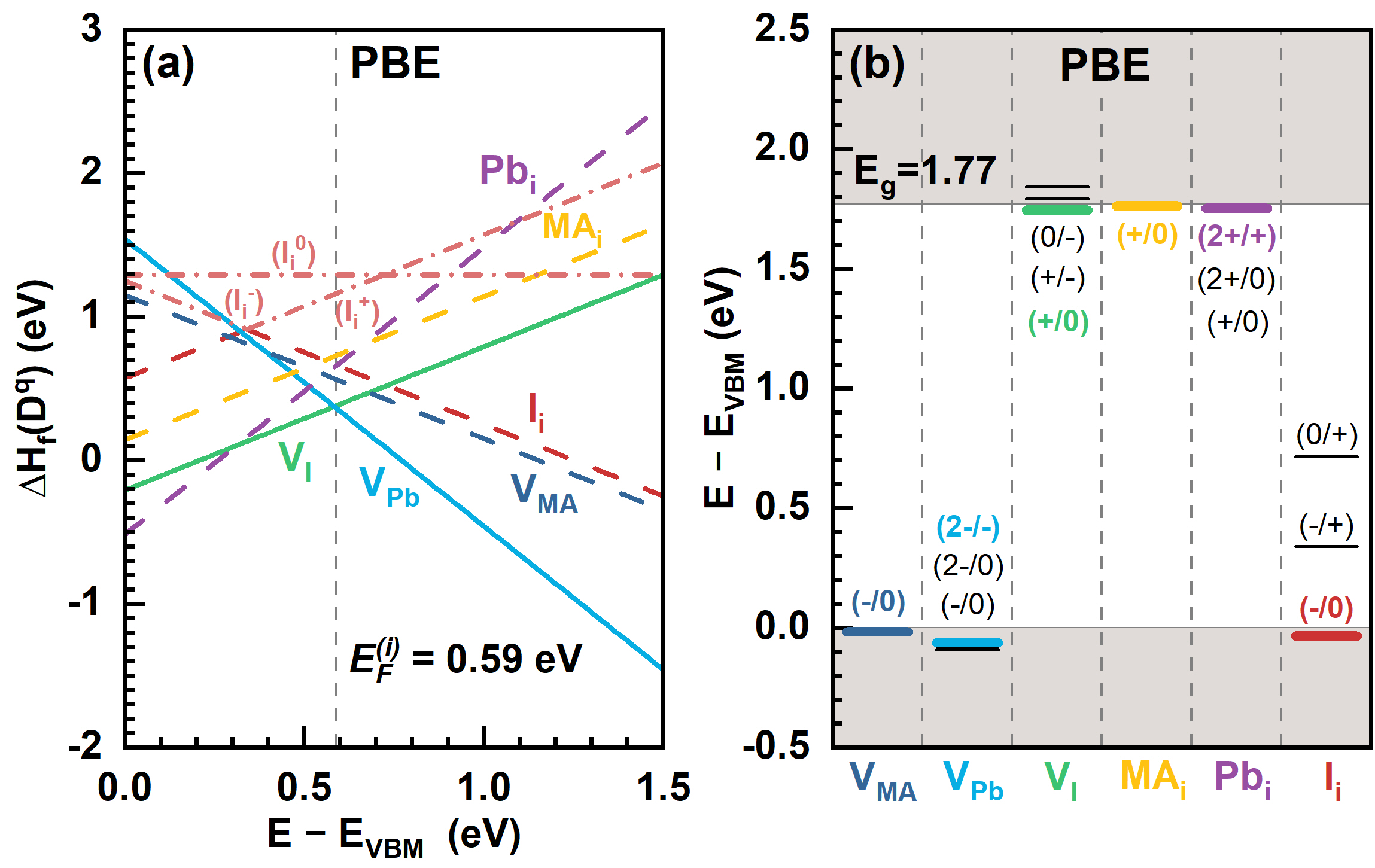}
    \caption{(a) Defect formation energies of vacancies and interstitials calculated with PBE; the solid lines give the most stable charged states; the dashed lines represent other charged states. (b) Charge state transition levels; the most important ones are indicated by colored lines. \label{figure:DFE_and_CSTL_of_PBE}}
\end{figure}

As PBE is the most widely employed functional, we will use it as reference for studying the defect thermodynamics and electronic properties. The calculated DFEs and CSTLs are shown in Fig. \ref{figure:DFE_and_CSTL_of_PBE}. At equilibrium conditions, as expressed by the charge neutrality condition, Eq. \ref{eq:charge_neutrality}, the intrinsic Fermi level $E_F^{(i)}=0.59$ eV. At these conditions, the most stable point defects are the vacancies $\mathrm{V_{Pb}}^{2-}$ and $\mathrm{V_I}^+$ with DFEs of 0.36 eV and 0.38 eV, respectively, which leads to equilibrium concentrations, Eq. \ref{eq:defect_concentration}, of $2.8\times 10^{15}$ $\mathrm{cm}^{-3}$ and $5.7\times 10^{15}$ $\mathrm{cm}^{-3}$ for these defects, respectively. 

The formation energies of the other elementary point defects $\mathrm{V_{MA}}^-$, $\mathrm{MA_i}^+$, $\mathrm{Pb_i}^{2+}$ and $\mathrm{I_i}^-$, are 0.2 to 0.4 eV higher, resulting in a lower equilibrium concentration by a factor of $10^3$ to $10^6$. PBE calculations thus support the classic picture of Schottky defects in ionic crystals, as represented by the oppositely charged vacancies $\mathrm{V_{Pb}}^{2-}$ and $\mathrm{V_I}^+$.

The CSTLs of these defects, calculated with the PBE functional, are shown in Fig. \ref{figure:DFE_and_CSTL_of_PBE}(b). One defect can have more than two charge states, and thus more than one transition level, where it is also possible that the charge changes by more than one unit $e$. However, starting from one stable charge state of a defect, only the transition levels representing a change of a single unit $\pm e$ are considered to be active. This is because the probability of capturing two or more holes/electrons simultaneously is very low \cite{Tress2017, Meggiolaro2018review}.

Marking the CSTLs in Fig. \ref{figure:DFE_and_CSTL_of_PBE}(b) where the charge state changes by more than $\pm e$ as inactive, then shows that the remaining levels are all within 10 $k_BT$ (0.26 eV at the room temperature) from the band edges, which qualifies them as shallow impurities \cite{Walle2004, Freysoldt2014, Park2018, Buin2014}.

\subsection{SOC, Van der Waals and meta-GGA functionals} \label{section:comparePBE}

Keeping the PBE results as a reference, we now turn to different functionals. The most obvious ingredients to add, are Van der Waals (vdW) interactions. The simplest approach uses parameterized semi-empirical expressions on top of the PBE functional, leading to the PBE-D3 or PBE-D3(BJ) functionals \cite{Grimme2010,Grimme2011}. An alternative approach defines explicit non-local terms representing vdW interactions. They can be incorporated in the density functional, as in the rev-vdW-DF2 functional, or added seamlessly to an existing functional, such as rVV10. Switching from a standard GGA to a more advanced meta-GGA functional, we also use the SCAN functional, without and with explicit vdW terms, the latter in the form of SCAN+rVV10. Spin-orbit coupling (SOC) can be added straightforwardly to the Hamiltonian, irrespective of the density functional.

\begin{figure}
    \includegraphics[width=1\columnwidth]{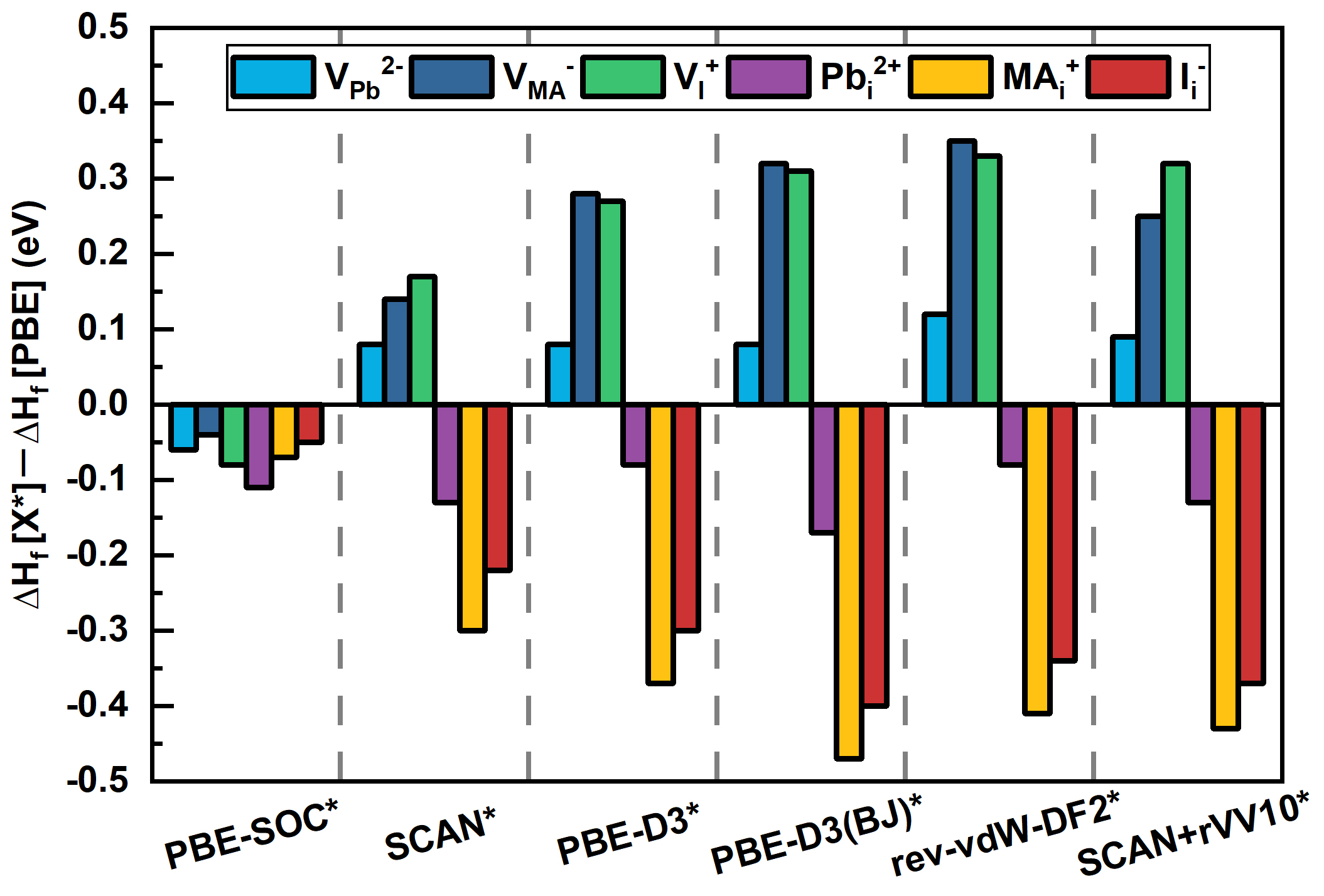}
    \caption{Differences between defect formation energies $\Delta H_f[X^*]$, calculated with different functionals for PBE optimized structures, and $\Delta H_f[\mathrm{PBE}]$ calculated with PBE. \label{figure:delta_DFE_pbe}}
\end{figure}

Previous defect studies on hybrid metal halide perovskites often adopt the strategy of applying vdW interactions, SOC, or hybrid functionals to the equilibrium structures optimized by PBE, in order to avoid the high computational costs associated with geometry optimization \cite{Du2015, Meggiolaro2018-iodine-chemistry, Meggiolaro2018review}. We perform two sets of calculations, a first set where the different functionals are used on the structures optimized by PBE, and a second set where all structures are optimized self-consistently with the respective functionals. This allows us to assess the effects on the DFEs of replacing the functional, and of geometry optimization separately. The results of the first set of calculations are discussed in this section, and those of the second set in the next section. The changes in the calculated DFEs if switching from the PBE functional to one of the other functionals (while keeping the geometries fixed), are shown in Fig. \ref{figure:delta_DFE_pbe}. 

This figure demonstrates that adding SOC, while holding on to the PBE functional, leads to relatively small changes in the DFEs. SOC decreases the DFEs of all elementary vacancy and interstitial point defects by $0.04$-$0.11$ eV. Moreover, switching to the SCAN functional, the differences between the DFEs calculated with and without SOC, are very similar to those obtained by PBE, as is illustrated by Table \ref{table:effect of SOC}. This indicates (i) that the effect of SOC on the DFEs is small, and (ii) that it can be added as an a posteriori correction. 

\begin{table}
    \caption{Differences of defect formation energies (eV) induced by adding SOC on top of PBE and SCAN, respectively. Here, single point calculations including SOC are based on equilibrium structures optimized by PBE and SCAN, respectively.}
    \label{table:effect of SOC}
    \begin{ruledtabular}
    \begin{tabular}{lcccccc}
        Functional & $\mathrm{V_{MA}}^-$ & $\mathrm{V_{Pb}}^{2-}$ & $\mathrm{V_{I}}^+$ & $\mathrm{MA_i}^+$ & $\mathrm{Pb_i}^{2+}$ & $\mathrm{I_i}^-$\\
        \hline
        PBE-SOC	& $-0.04$ & $-0.06$ & $-0.08$ & $-0.07$ & $-0.11$ & $-0.05$\\
        SCAN-SOC & $-0.03$ & $-0.04$ & $-0.05$ & $-0.07$ & $-0.12$ & $-0.02$\\
    \end{tabular}
    \end{ruledtabular}
\end{table}

In contrast, adding vdW corrections to PBE (PBE-D3, PBE-D3(BJ)), or using a vdW functional (rev-vdW-DF2), brings significant changes of up to $\sim 0.4$ eV to the DFEs, see Fig. \ref{figure:delta_DFE_pbe}. All interstitials are stabilized, whereas all vacancies are destabilized. Van der Waals interactions generally increase the bonding strength between atoms, molecules, and ions, which makes it easier to insert one of these in an existing lattice, and more difficult to extract one. Comparing the effect of adding vdW interactions on the DFEs of specific defects, one notices that the DFE changes of Pb interstitials and vacancies are only $\sim 0.1$ eV, whereas MA and iodine vacancies and interstitials are (de)stabilized by 0.3 to 0.4 eV. 

The SCAN functional is reported to capture intermediate-range vdW-type interactions \cite{Sun2016}, which is confirmed by Fig. \ref{figure:delta_DFE_pbe}. Compared to PBE, SCAN corrects the DFEs in the same direction as adding explicit vdW terms does, i.e., it stabilizes interstitials and destabilizes vacancies, albeit quantitatively to a lesser extend. Adding on top of this explicit long-range vdW interactions, as in SCAN+rVV10, then enhances these corrections. 

\subsection{Structure optimizations}\label{section:compareopt}

\begin{figure}
    \includegraphics[width=1\columnwidth]{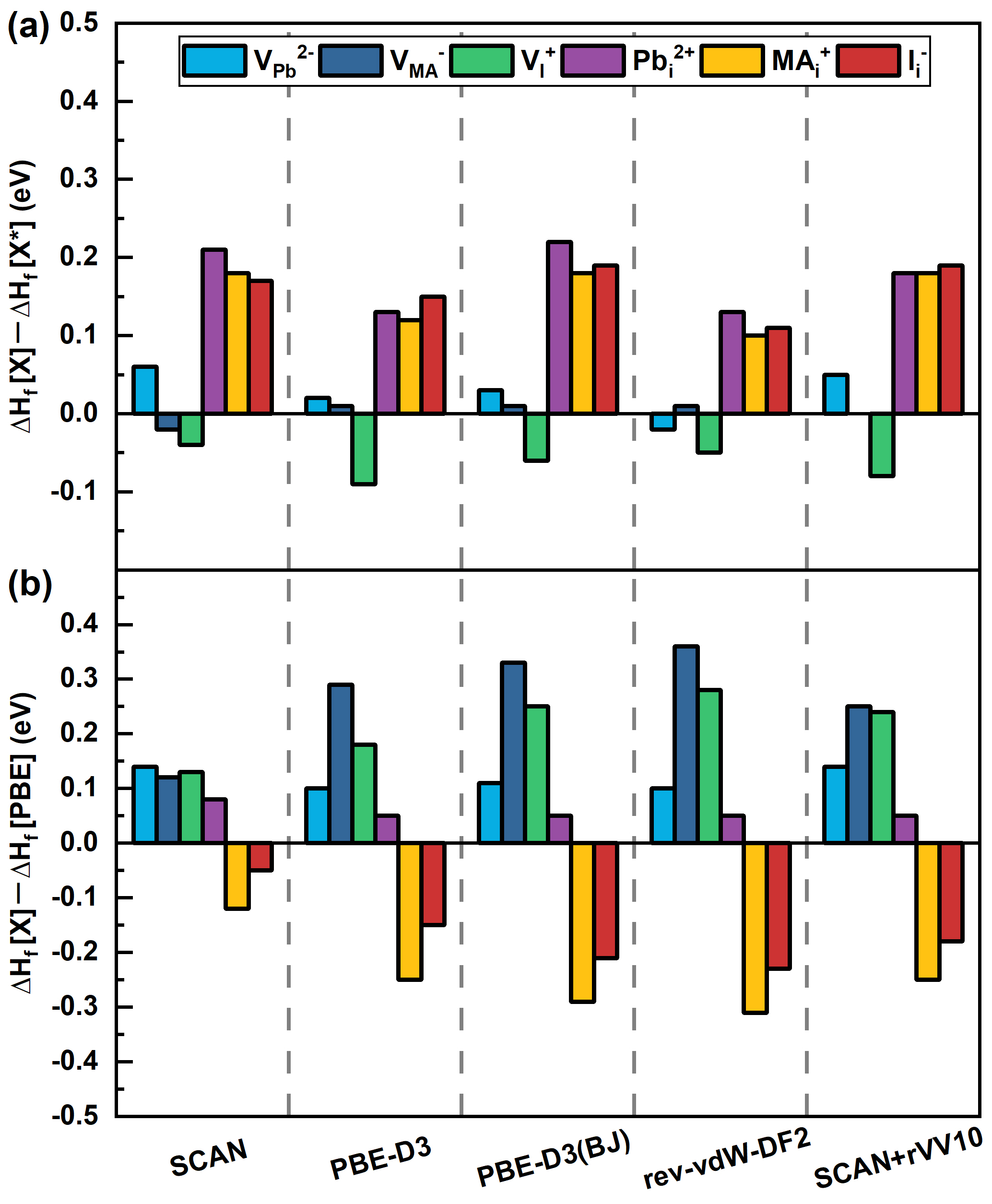}
    \caption{(a) Differences between defect formation energies $\Delta H_f[X]$, calculated with different functionals with optimized structures, and $\Delta H_f[X^*]$ with PBE structures. (b) Differences between $\Delta H_f[X]$ and $\Delta H_f[\mathrm{PBE}]$ (DFEs calculated with PBE). \label{figure:effect_of_opt}}
\end{figure}

As a next step we consider optimization of the structure with each functional. Figure \ref{figure:effect_of_opt}(a) shows the changes in the DFEs upon structure optimization, using the PBE-optimized structures as a reference. The DFEs of vacancies are only mildly affected; $\mathrm{V_{I}}^+$ becomes more stable by $\lesssim 0.1$ eV, $\mathrm{V_{Pb}}^{2-}$ becomes less stable by $\lesssim 0.05$ eV, and $\mathrm{V_{MA}}^-$ is barely affected at all. In contrast, the DFEs of interstitials become significantly larger upon structure optimization by $\sim 0.2$ eV. 

\begin{figure}
    \includegraphics[width=1\columnwidth]{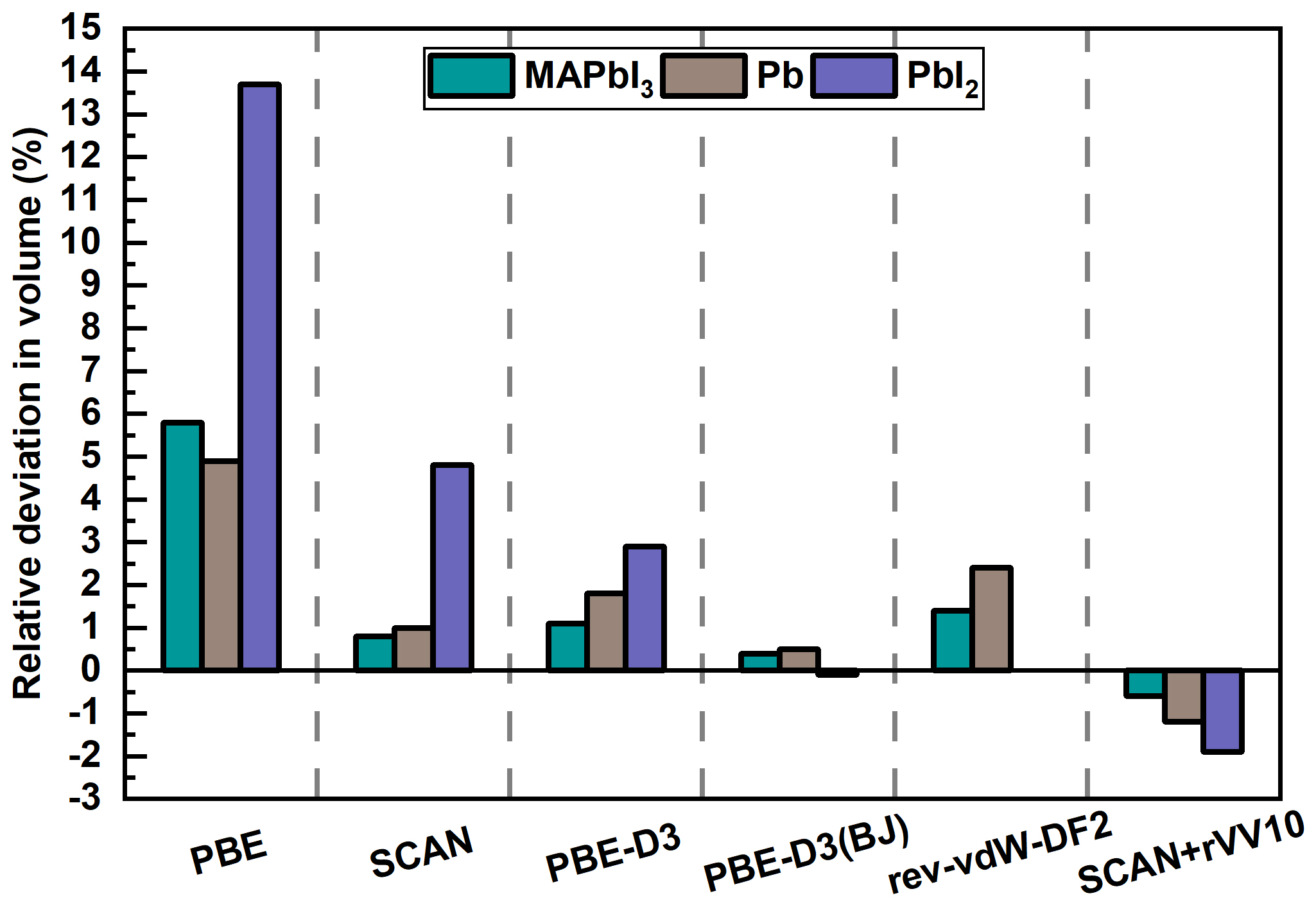}
    \caption{Deviations of equilibrium volumes of MAPbI$_3$, Pb metal and PbI$_2$, calculated by different functionals, relative to the experimental volumes \cite{Stoumpos2013, Vishan1976, Flahaut2006}. \label{figure:volume_deviation}}
\end{figure}

We correlate this markedly asymmetric behavior of vacancies and interstitials to differences in lattice volumes calculated with PBE, or with any of the other functionals. The deviation from the experimental value \cite{Stoumpos2013} of the optimized volume of MAPbI$_3$ is shown in Fig. \ref{figure:volume_deviation}. For comparison, also shown in this figure are the corresponding deviations for PbI$_2$ and Pb metal \cite{Vishan1976, Flahaut2006}. Immediately obvious is that PBE significantly overestimates the volume of MAPbI$_3$ by $\sim 6$\%. In this respect, the other functionals perform much better, as they give a notable improvement to volumes within 0.5-1.5\% of the experimental value. It should be noticed that the differences in volume are not accompanied by large changes in the bond distance or strength of the Pb-I bonds. Instead, the volume changes are mainly incorporated by changes of I-Pb-I angles, which leads to tilting of PbI$_6$ octahedra.

Similarly, PBE overestimates the volume of Pb metal by $\sim 5$\%, and the other functionals bring this deviation down to 0.5-2\%. PbI$_2$ is somewhat special, as it is a layered compound, which requires inclusion of vdW terms to capture the interaction between the two-dimensional PbI$_2$ layers. Not surprisingly then, PBE overestimates the PbI$_2$ volume by $\sim 14$\%, as it lacks these terms. Including D3 vdW terms corrects this to $\sim 3$\%, whereas the D3(BJ) vdW terms and the rev-vdW-DF2 functional reproduce the experimental value more or less exactly. The PBE error is more than halved by the SCAN functional regarding the volume of PbI$_2$, confirming the notion that SCAN captures some of the intermediate-range vdW attraction \cite{Sun2016}. However, capturing the longer range interactions requires adding explicit long-range vdW terms, such as rVV10. The SCAN+rVV10 functional overshoots the correction a little, and gives a volume of PbI$_2$ that is $\sim 2$\% too small. More details about the optimized lattice parameters can be found in the Supplemental information, Tables S1 and S2.

Relating these volume changes to the DFEs, one observes that, compared to PBE, all other functionals in Fig. \ref{figure:volume_deviation} give a markedly smaller volume for the MAPbI$_3$ lattice. One might expect that it becomes more difficult to insert an additional cation or anion in a smaller lattice, which explains why the DFEs of interstitials increases if optimizing the lattice starting from the PBE volume. The vacancies are generally much less affected by volume changes, as they can more easily be compensated by changes in I-Pb-I angles, and tilting of octahedra.

The changes in the DFEs, calculated via self-consistent structure optimization with the different functionals, with respect to the PBE-calculated results, are shown in Fig. \ref{figure:effect_of_opt}(b). They show a similar qualitative trend as the results shown in Fig. \ref{figure:delta_DFE_pbe}, which were obtained without structure optimization, i.e., vacancies are less stable and interstitials are generally more stable. However, in particular for interstitials the size of these changes is notably smaller if structure optimization is included. The change for $\mathrm{Pb_i}^{2+}$ becomes $\lesssim 0.05$ eV, whereas $\mathrm{I_i}^{-}$ becomes more stable by 0.05-0.25 eV. Remarkably, the largest effect is for $\mathrm{MA_i}^{+}$, which is stabilized by 0.1-0.3 eV.

In general, SCAN gives DFEs that are within 0.15 eV of PBE, although the changes for vacancies and interstitials tend to be of opposite sign. Including vdW interactions enlarges these changes up to 0.3 eV, while maintaining the same trend. This number is still notably smaller than what is obtained without structure optimization, which emphasizes the importance of performing structure optimization self-consistently if switching the functional. 

\subsection{Defect formation energies and concentrations}

\begin{table*}
    \caption{Formation energies $\Delta H_f$ (eV) and concentrations $c$ (cm$^{-3}$) of different defects in MAPbI$_3$ calculated by different functionals, including self-consistent geometry optimization. The asterisk * represents single point calculations including SOC, based on PBE optimized structures.}
    \label{table:Concentration_of_all_defects}
    \begin{ruledtabular}
    \begin{tabular}{lllllll}
        Functional & $\mathrm{V_{MA}}^-$ & $\mathrm{V_{Pb}}^{2-}$ & $\mathrm{V_{I}}^+$ & $\mathrm{MA_i}^+$ & $\mathrm{Pb_i}^{2+}$ & $\mathrm{I_i}^-$\\
        \hline
\multicolumn{7}{l}{Defect formation energy $\Delta H_f$ (eV)}\\
PBE & 0.56 & 0.36 & 0.38 & 0.72 & 0.65 & 0.67\\
PBE-SOC* & 0.52 & 0.30 & 0.31 & 0.67 & 0.55 & 0.61\\
SCAN & 0.68 & 0.50 & 0.51 & 0.61 & 0.74 & 0.61\\
PBE-D3 & 0.86 & 0.47 & 0.55 & 0.48 & 0.70 & 0.51\\
PBE-D3(BJ) & 0.89 & 0.47 & 0.63 & 0.44 & 0.71 & 0.45\\
rev-vdW-DF2 & 0.91 & 0.45 & 0.66 & 0.42 & 0.72 & 0.43\\
SCAN+rVV10 & 0.81 & 0.50 & 0.61 & 0.47 & 0.71 & 0.48\\
\multicolumn{7}{l}{Defect concentration $c$ (cm$^{-3}$)}\\ 
PBE & 1.27$\times 10^{12}$ & 2.83$\times 10^{15}$ & 5.66$\times 10^{15}$ & 7.74$\times 10^{9}$ & 1.17$\times 10^{11}$ & 7.58$\times 10^{10}$\\
PBE-SOC* & 7.66$\times 10^{12}$ & 3.93$\times 10^{16}$ & 7.87$\times 10^{16}$ & 7.67$\times 10^{10}$ & 6.16$\times 10^{12}$ & 6.58$\times 10^{11}$\\
SCAN & 1.39$\times 10^{10}$ & 1.45$\times 10^{13}$ & 2.89$\times 10^{13}$ & 6.46$\times 10^{11}$ & 4.28$\times 10^{9}$ & 6.01$\times 10^{11}$\\
PBE-D3 & 1.58$\times 10^{7}$ & 4.63$\times 10^{13}$ & 6.44$\times 10^{12}$ & 1.18$\times 10^{14}$ & 2.22$\times 10^{10}$ & 3.19$\times 10^{13}$\\
PBE-D3(BJ) & 3.95$\times 10^{6}$ & 4.57$\times 10^{13}$ & 3.41$\times 10^{11}$ & 4.12$\times 10^{14}$ & 1.37$\times 10^{10}$ & 3.21$\times 10^{14}$\\
rev-vdW-DF2 & 1.89$\times 10^{6}$ & 1.18$\times 10^{14}$ & 8.84$\times 10^{10}$ & 9.64$\times 10^{14}$ & 1.08$\times 10^{10}$ & 7.28$\times 10^{14}$\\
SCAN+rVV10 & 8.68$\times 10^{7}$ & 1.44$\times 10^{13}$ & 6.10$\times 10^{11}$ & 1.37$\times 10^{14}$ & 1.59$\times 10^{10}$ & 1.09$\times 10^{14}$\\
    \end{tabular}
    \end{ruledtabular}
\end{table*}

The DFEs under intrinsic conditions and the equilibrium concentrations of the elementary point defects in MAPbI$_3$, calculated with the different functionals including self-consistent structure optimization, are given in Fig. \ref{figure:DFE_opt_combine}. For comparison, the DFEs calculated by different functionals based on PBE-optimized structures are shown in Figure S5. As we have argued in the previous section, in order to obtain accurate results, it is important to perform a self-consistent structure optimization for each functional.

As discussed in Sec. \ref{section:PBEresults}, the PBE functional yields the lowest DFEs for the standard Schottky defects $\mathrm{V_{Pb}}^{2-}$ and $\mathrm{V_I}^+$, whereas the DFEs of the other point defects are significantly larger. In terms of equilibrium concentrations, this means that $\mathrm{V_{Pb}}^{2-}$ and $\mathrm{V_I}^+$ are the dominant effects by far. Adding SOC gives little change; all DFEs are slightly reduced, increasing equilibrium concentrations by roughly an order of magnitude, but maintaining approximately the ratios between different defects. As argued in Sec. \ref{section:comparePBE}, adding SOC has the same effect also for other functionals. 

Using the SCAN functional, one observes that the DFEs of the Schottky defects $\mathrm{V_{Pb}}^{2-}$ and $\mathrm{V_I}^+$ still are the smallest, but they are somewhat larger than those obtained with PBE, so that their equilibrium concentration is roughly two orders of magnitude smaller. In addition, the SCAN DFEs of the interstitials $\mathrm{MA_i}^+$ and $\mathrm{I_i}^-$, although still larger than the vacancy DFEs, have become smaller, such that their equilibrium concentrations are within a factor of fifty or so from that of the vacancies.

Adding vdW interactions (PBE-D3, PBE-D3(BJ), rev-vdW-DF2, SCAN+rVV10) enforces this trend. In fact, the DFEs of the interstitials $\mathrm{MA_i}^+$ and $\mathrm{I_i}^-$ are now the smallest of all points defects, where only the DFE of the lead vacancy, $\mathrm{V_{Pb}}^{2-}$, has a similar magnitude. This implies that $\mathrm{MA_i}^+$ and $\mathrm{I_i}^-$ have the largest equilibrium concentrations, with the $\mathrm{V_{Pb}}^{2-}$ concentration being of the same order of magnitude. The equilibrium concentration of the iodine vacancy $\mathrm{V_{I}}^{+}$ is typically two to four orders of magnitude smaller, and the concentrations of $\mathrm{Pb_i}^{2+}$ and $\mathrm{V_{MA}}^{-}$ are even smaller. 

There are small differences between the results calculated with the different functionals that include vdW interactions, but the DFEs calculated with PBE-D3(BJ), rev-vdW-DF2 and SCAN+rVV10 are within $\sim 0.05$ eV of one another, and even PBE-D3 gives values that are quite close. The calculated densities of dominant defects fall in the range $10^{13}$-$10^{15}$ cm$^{-3}$, which is consistent with the experimentally reported range of values \cite{Adinolfi2016, Leijtens2016, Chen2016a, DeQuilettes2015, Blancon2016, Samiee2014}.

\subsection{Changing the Fermi level}

\begin{figure}
    \includegraphics[width=1\columnwidth]{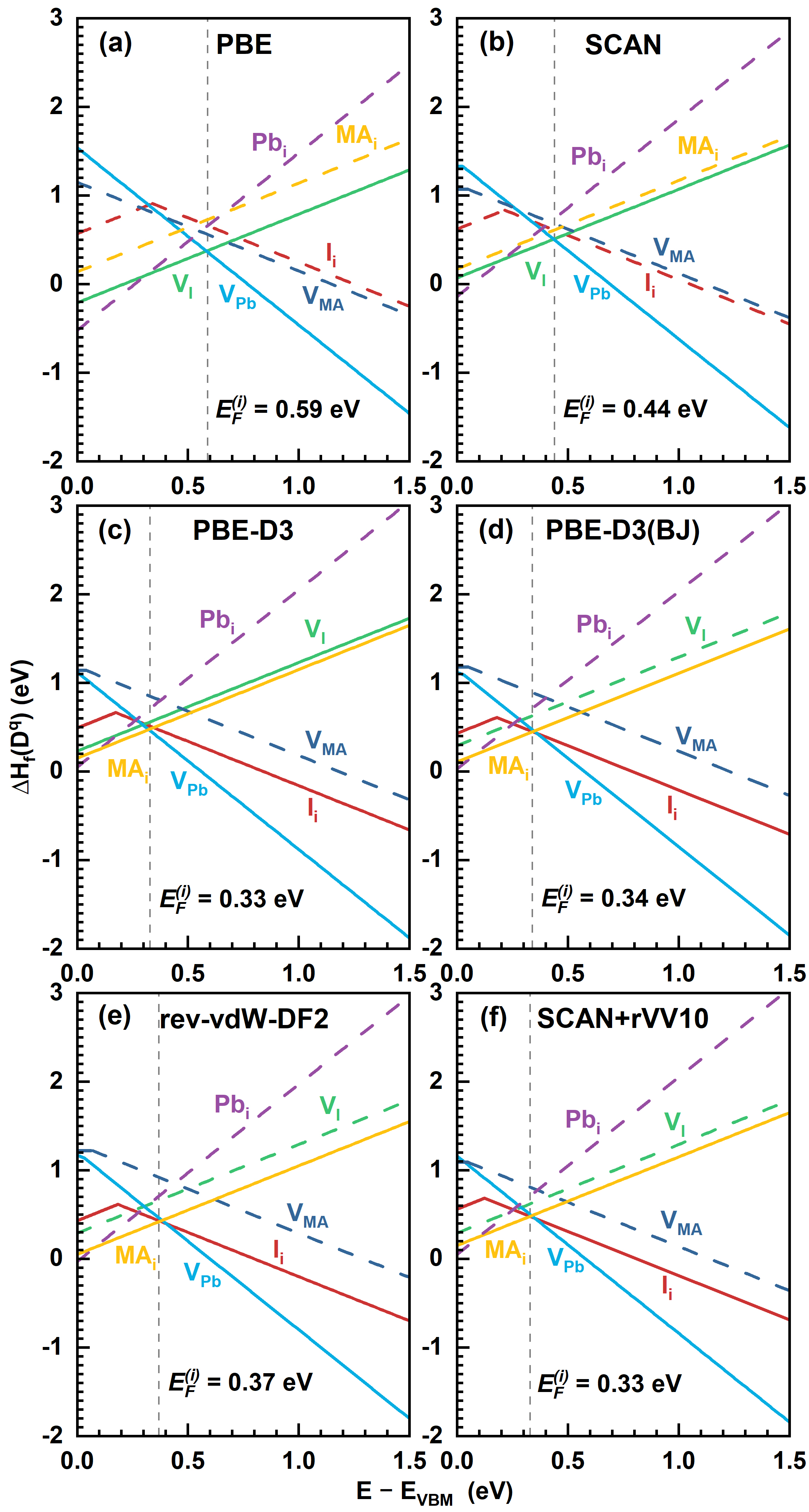}
    \caption{Defect formation energies calculated by different functionals including structure optimization. \label{figure:DFE_opt_combine}}
\end{figure}

\begin{figure}
    \includegraphics[width=1\columnwidth]{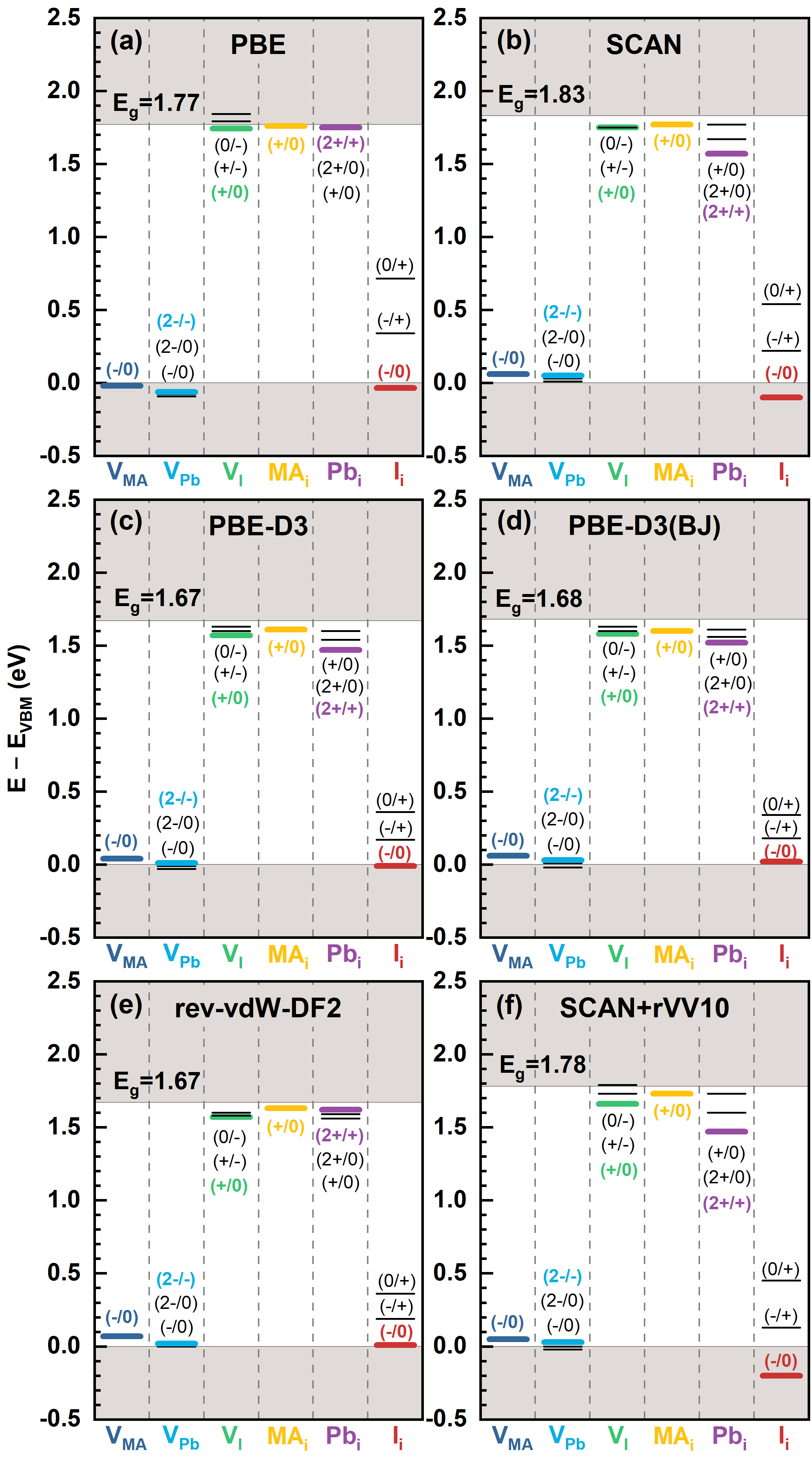}
    \caption{Charge state transition levels calculated by different functionals including structure optimization. \label{figure:CSTL_opt_combine}}
\end{figure}

Figure \ref{figure:DFE_opt_combine} displays the DFEs as a function of the position of the Fermi level $E_F$, calculated with full structure optimization. Results obtained with PBE-optimized structures are shown in the Supplemental Information, Figs. S5 and S6. Figure \ref{figure:DFE_opt_combine} demonstrates the differences between the DFEs calculated with different functionals, discussed in the previous section. In particular, at the intrinsic Fermi levels (the dashed vertical lines in Fig. \ref{figure:DFE_opt_combine}) the interstitials $\mathrm{MA_i}^+$ and $\mathrm{I_i}^-$ and the vacancy $\mathrm{V_{Pb}}^{2-}$ have the smallest DFE for the functionals that include vdW interactions, whereas without vdW interactions the vacancies $\mathrm{V_{I}}^{+}$ and $\mathrm{V_{Pb}}^{2-}$ have the smaller DFE. The figure also illustrates that most defects stay in the same charge states over a large range of $E_F$. Only the iodine interstitial, $\mathrm{I_i}$, shows a change from $+$ to $-$ charge at $E_F$ well inside the gap (varying from 0.1 eV above the VBM for SCAN+rVV10 to 0.3 eV above VBM for PBE).

Using such results, the CSTLs can be determined for all point defects for all functionals used. The results are shown in Fig. \ref{figure:CSTL_opt_combine}. One can observe that the CSTLs of most defects are close to the band edges for all functionals, implying that they are shallow traps, which is consistent with the common notion that electronically MAPbI$_3$ is defect tolerant \cite{Yin2014, Walsh2015, Steirer2016}. One exception is $\mathrm{I_i}$, where the CSTLs $(-/+)$ and $(0/+)$ are well inside the gap for all functionals. However, the $(-/+)$ CSTL is inactive under device operating conditions, as it requires the simultaneous capture of two charge carriers of the same type, which is highly unlikely. Meggiolaro \emph{et al.} have argued that the $(0/+)$ CSTL is also inactive under these conditions, as it involves a considerable structural change accompanied by an energy barrier \cite{Meggiolaro2018-iodine-chemistry}. This remains true for all functionals tested here.
 
A second exception is (2+/+) CSTL of $\mathrm{Pb_i}$ calculated with SCAN or SCAN+rVV10, at 0.26 eV and 0.31 eV below the CBM, respectively, which is somewhat too deep to be termed a shallow trap. However, the concentration of $\mathrm{Pb_i}$ defects will be extremely small due to its large DFE.  

\section{Summary and conclusions} \label{section: conclusions}
We have studied the intrinsic point defects, vacancies and interstitials, in the archetype organometal hybrid perovskite, MAPbI$_3$, by means of DFT calculations, employing a wide range of functionals. Although such defects do not seem to hamper the electronic operation of perovskite solar cells, they are vital in understanding the instabilities and degradation mechanisms of perovskite materials. Identifying the dominant defects from DFT calculations is hampered by the fact that different functionals give different results. Using the standard PBE (GGA) functional as a starting point, we have systematically investigated the effects of adding Van der Waals (vdW) interactions, either in simple parametrized form (D3, D3(BJ)), or as non-local functional (rev-vdW-DF2, rVV10), switching to meta-GGA (SCAN), as well as including spin-orbit coupling (SOC).

Besides on the parent material MAPbI$_3$, defect formations energies (DFEs) depend on the chemical potentials of the elements involved, and in case of charged defects, on the position of the Fermi level, which draws in inaccuracies in thermodynamic (formation enthalpies), as well as electronic properties (band positions), caused by DFT functionals. Focusing on intrinsic thermodynamic conditions, we show that the DFEs of MAPbI$_3$ are relatively insensitive to these inaccuracies. The same will hold for other organometal hybrid perovskites.  

Nevertheless, different functionals present different defects as most common. Whereas PBE predicts the standard Schottky defects, iodine and lead vacancies, $\mathrm{V_I}^{+}$ and $\mathrm{V_{Pb}}^{2-}$, to be dominant, including vdW interactions favors the iodine and methylammonium interstitials, $\mathrm{I_i}^-$ and $\mathrm{MA_i}^+$, besides lead vacancies, $\mathrm{V_{Pb}}^{2-}$. In general, vdW interactions stabilize interstitials, and destabilize vacancies.

In addition, we conclude that self-consistent structural optimization is important in order to obtain accurate DFEs. We correlate this outcome to a difference in equilibrium volume found for MAPbI$_3$ by the different functionals. PBE in particular gives a volume that is too large. If one adds vdW interactions without reoptimizing the structure and volume, one overestimates the stability of interstitials in particular.  

The meta-GGA (SCAN) functional performs overall better than GGA (PBE), evidenced by the consistent improvements in lattice parameters of the relevant perovskite (MAPbI$_3$) and precursor (PbI$_2$). However, regarding DFEs, SCAN only captures part of the non-bonding interactions between the organic cations and the inorganic iodine and lead ions, and we suggest to include long-range vdW interactions in the form of SCAN+rVV10. Adding vdW terms to PBE, D3(BJ) in particular, or using the vdW functional rev-vdW-DF2 also gives decent DFEs. However, given the proven versatility of SCAN to accurately describe a large variety of compounds and molecules in different bonding configurations, we express a preference for the SCAN+rVV10 functional.
 
In contrast, the inclusion of SOC gives only relatively small and consistent corrections to the DFEs, which can be added straightforwardly as a post-correction.

\begin{acknowledgments}
H. Xue acknowledges the funding from the China Scholarship Council (CSC). S. Tao acknowledges funding by the Computational Sciences for Energy Research (CSER) tenure track program of Shell and NWO (Project number 15CST04-2) and the NWO START-UP grant from the Netherlands.
\end{acknowledgments}

\bibliography{Manuscript_comparison_of_functionals}

\bibliographystyle{apsrev4-2}

\end{document}


\title{Supplemental Material\\ First-principles Calculations of Defects in Metal Halide Perovskites: \\ a Performance Comparison of Density Functionals}

\author{Haibo Xue}
\affiliation{Materials Simulation and Modelling, Department of Applied Physics, Eindhoven University of Technology, P.O. Box 513, 5600MB Eindhoven, the Netherlands.}
\affiliation{Center for Computational Energy Research, Department of Applied Physics, Eindhoven University of Technology, P.O. Box 513, 5600MB Eindhoven, the Netherlands.}

\author{Geert Brocks}
\affiliation{Materials Simulation and Modelling, Department of Applied Physics, Eindhoven University of Technology, P.O. Box 513, 5600MB Eindhoven, the Netherlands.}
\affiliation{Center for Computational Energy Research, Department of Applied Physics, Eindhoven University of Technology, P.O. Box 513, 5600MB Eindhoven, the Netherlands.}
\affiliation{Computational Materials Science, Faculty of Science and Technology and MESA+ Institute for Nanotechnology, University of Twente, P.O. Box 217, 7500AE Enschede, the Netherlands.}

\author{Shuxia Tao}
\email{s.x.tao@tue.nl}
\affiliation{Materials Simulation and Modelling, Department of Applied Physics, Eindhoven University of Technology, P.O. Box 513, 5600MB Eindhoven, the Netherlands.}
\affiliation{Center for Computational Energy Research, Department of Applied Physics, Eindhoven University of Technology, P.O. Box 513, 5600MB Eindhoven, the Netherlands.}

\maketitle

\tableofcontents

\clearpage

\section{Optimized Lattice Parameters and Volumes}
\subsection{MAPbI$_3$}
The optimized lattice parameters of the MAPbI$_3$ tetragonal unit cell, as well as the volumes per formula unit $V_{f.u.}$, calculated by different functionals, are summarized in Table \ref{tab: volume of a formula unit}. 

\begin{table}[h!]
    \caption{Optimized lattice parameters of the tetragonal unit cell, and the volume per formula unit of MAPbI$_3$, calculated using different functionals.}
    \label{tab: volume of a formula unit}
    \centering
    \begin{ruledtabular}    
    \begin{tabular}{lcccc}
        \multirow{2}{*}{Functional} & \multicolumn{2}{c}{Lattice parameters} & \multicolumn{1}{c}{Volume per formula unit} \\
        \cline{2-3} \cline{4-4}
         & $a=b$ (\AA) & c (\AA) & $V_{f.u.}$ (\AA$^3)$ \\
        \hline
        Experiment \cite{Stoumpos2013} & 8.849 & 12.642 & 247.48 \\
        PBE & 9.016 & 12.881 &  261.77  \\
        SCAN & 8.873 & 12.676 &  249.50 \\
        PBE-D3 & 8.880 & 12.687 &  250.11	 \\
        PBE-D3(BJ) & 8.861 & 12.659 &  248.49	 \\
        rev-vdW-DF2 & 8.889 & 12.699 &  250.85 \\
        SCAN+rVV10 & 8.832 & 12.617 &  246.04 \\
    \end{tabular}
    \end{ruledtabular}
\end{table}

\newpage

\subsection{PbI$_2$}
\begin{figure} [h]
    \centering
    \includegraphics[width=0.7\columnwidth]{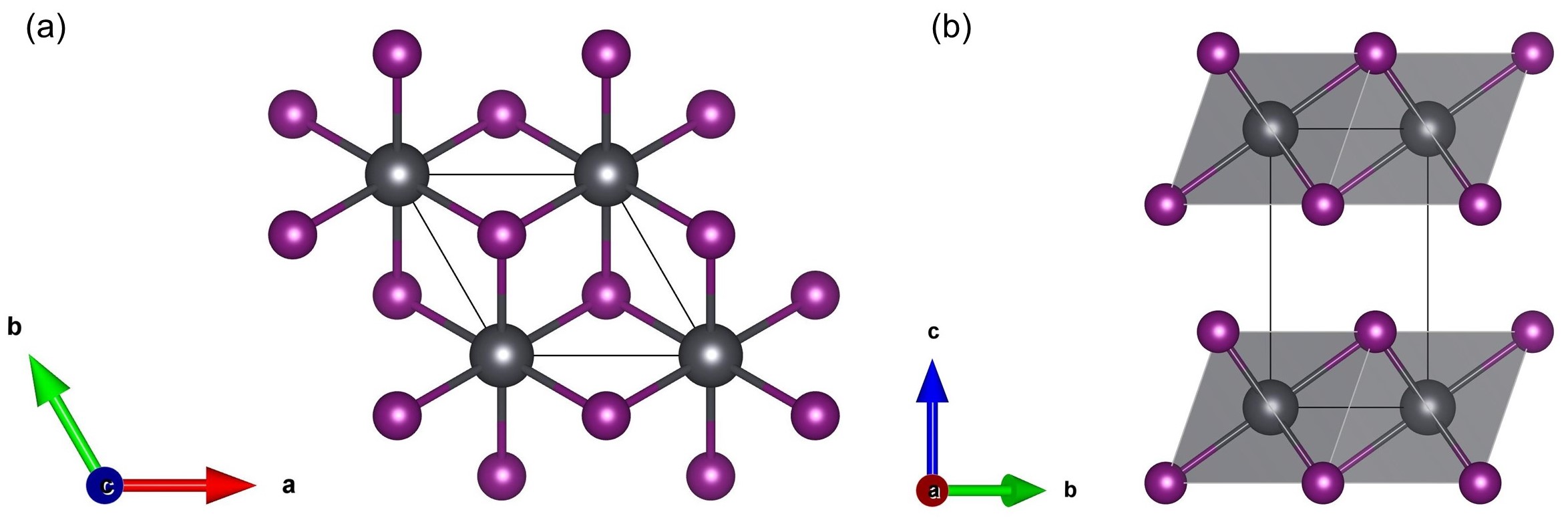}
    \caption{Top (a) and side (b) views of the layered structure of PbI$_2$.}
    \label{fig: PbI2}
\end{figure}

\begin{table}[h!]
    \caption{Deviation in lattice parameters of PbI$_2$. The lattice parameter $c$ is the inter-layer distance of PbI$_2$.}
    \label{tab: lattice parameters of PbI2}
    \centering
    \begin{ruledtabular}    
    \begin{tabular}{lcccc}
        \multirow{2}{*}{Functional} & \multicolumn{2}{c}{Lattice parameters} & \multicolumn{2}{c}{Deviation} \\
        \cline{2-3} \cline{4-5}
         & $a=b$ (\AA) & c (\AA) & $\Delta a,b$ (\%) & $\Delta c$ (\%) \\
        \hline
        Experiment \cite{Flahaut2006} & 4.558 & 6.986 & 0.0 & 0.0 \\
        PBE & 4.660 & 7.634 & 2.2 & 9.3 \\
        SCAN & 4.603 & 7.212 &  1.0 & 3.2 \\
        PBE-D3 & 4.610 & 7.063 & 1.1 & 1.1	 \\
        PBE-D3(BJ) & 4.563 & 6.999 & 0.1 & 0.2 \\
        rev-vdW-DF2 & 4.563 & 6.999 & 0.1 & 0.2 \\
        SCAN+rVV10 & 4.534 & 6.956 & -0.5 & -0.4 \\
    \end{tabular}
    \end{ruledtabular}
\end{table}

PbI$_2$ has a layered structure, where atoms are strongly bonded within two-dimensional layers, and the bonding between these 2D layers is much weaker (Fig. \ref{fig: PbI2}). In the main text we show that DFT functionals that include vdW terms, yield a much better equilibrium volume for PbI$_2$, than functionals without vdW terms. Table \ref{tab: lattice parameters of PbI2} shows the the lattice parameters of PbI$_2$, calculated with different functionals.

It can be seen that the functionals without vdW terms describe the inter-layer distance much worse than the in-plane lattice constants, as the $c$-axis is most connected to the inter-layer bonding [Fig. \ref{fig: PbI2}(b)]. PBE gives the length of the $c$-axis that is $\gtrsim 9$\% too large, whereas the in-plane lattice constants $a,b$ are less affected, with $\gtrsim 2$\% too large. SCAN corrects the errors to $\gtrsim 3$\% and 1\%, respectively. 

In contrast, the functionals with vdW terms significantly reduce the error in the inter-layer distance. And what is somewhat surprising, is that when adding vdW interactions, the in-plane lattice constants $a,b$ are affected in a very similar way. PBE-D3 approximately reduce the errors to $\gtrsim 1$\%, whereas PBE-D3(BJ) and rev-vdW-DF2 come within $\lesssim 0.2$\% of the experimental values. SCAN+rVV10 gives lattice parameters that are $\lesssim 0.5$\% too small.

\clearpage

\section{Defective Structures and Energies} \label{appendix: defective structures}

\subsection{Search strategy and energy optimization}
\begin{figure}[h]
    \includegraphics[width=0.7\textwidth]{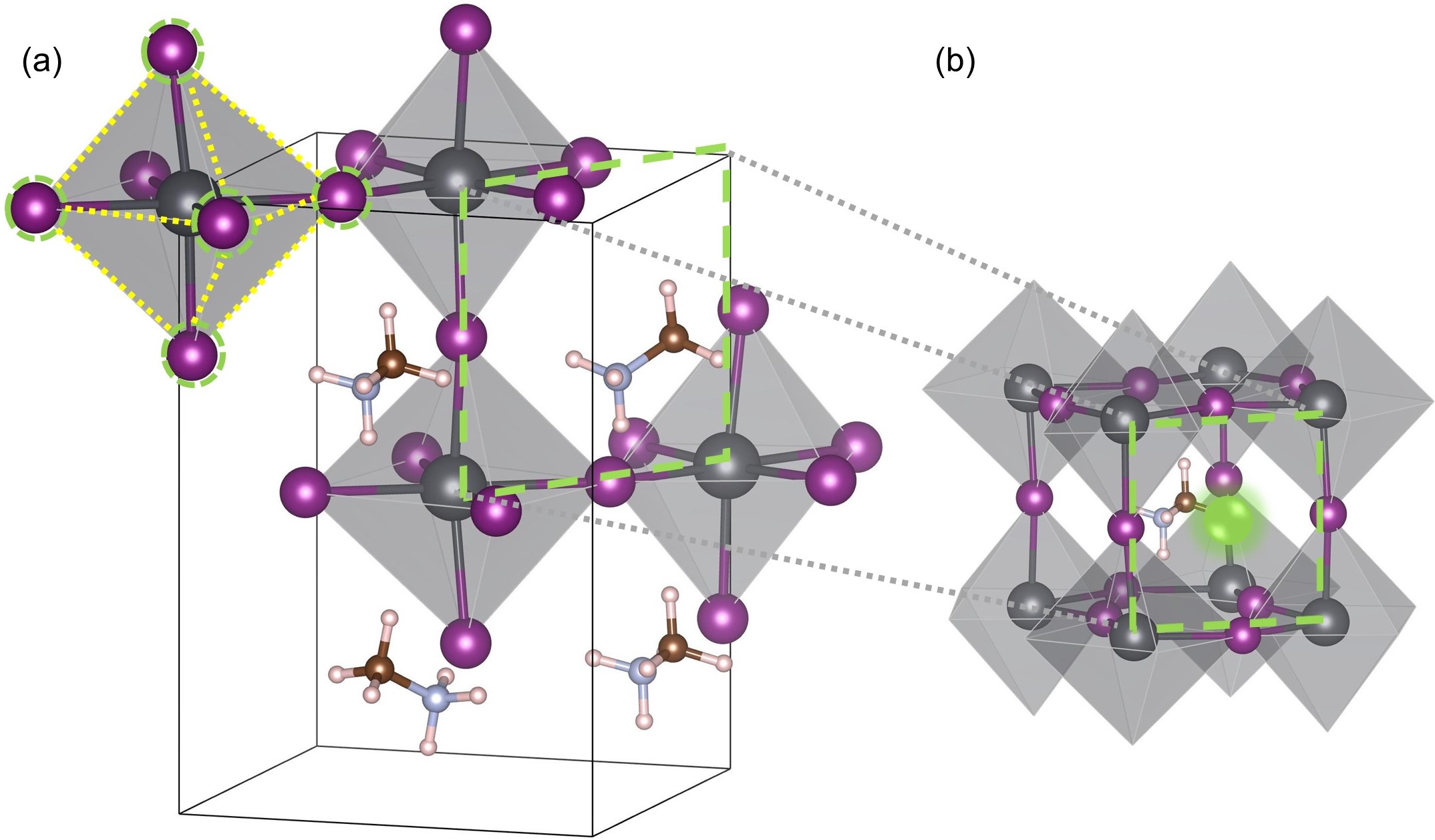}
    \caption{(a) Tetragonal unit cell of MAPbI$_3$, containing four MAPbI$_3$ formula units; one of the PbI$_6$ octahedra is highlighted; (b) schematic structure of the cube surrounding a MA ion; the green sphere indicates the center of a cube face.} \label{figure: scheme of creating defects}
\end{figure}

\begin{figure}
    \includegraphics[width=0.6\textwidth]{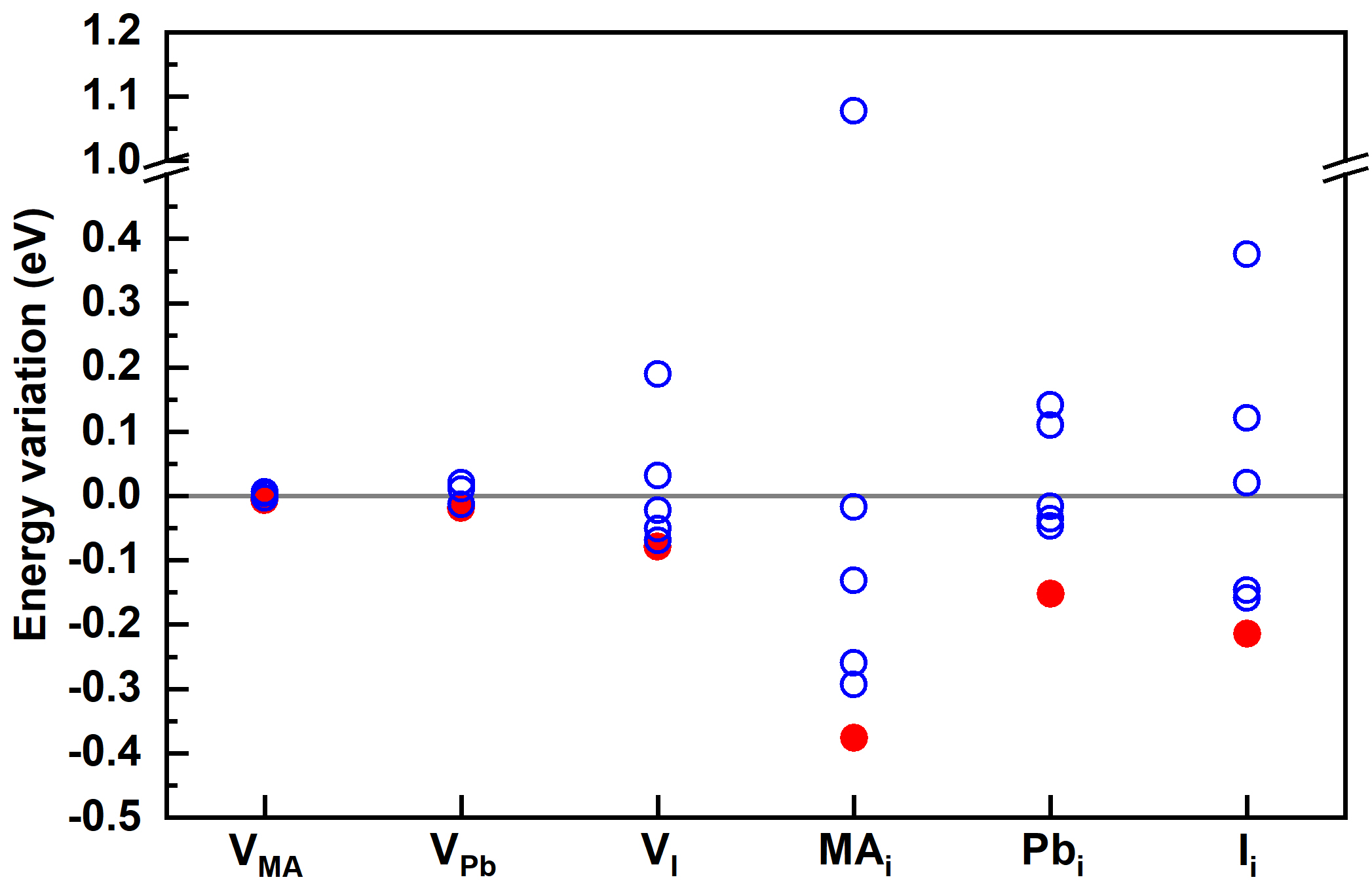}
    \caption{Total energies of defective structures, obtained with different starting configurations (see text). The energies shown are with respect to their average. The red dots indicate the lowest energies,  which we use in the main text to calculate the defect formation energies. \label{figure: energy_spread}}
\end{figure}

\begin{figure} 
    \includegraphics[width=0.8\textwidth]{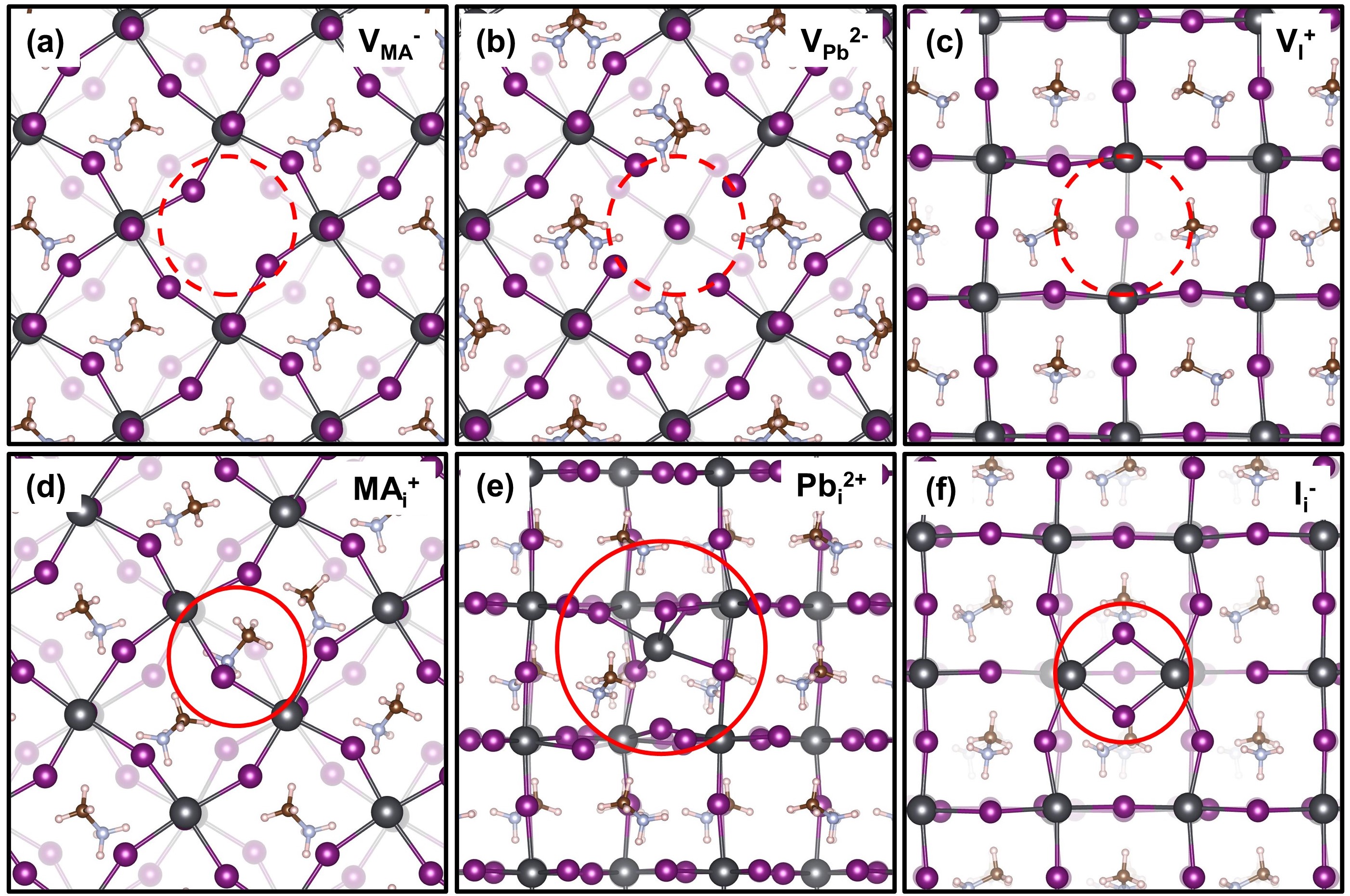}
    \caption{Optimized structures of different defects in their most stable charge states. From a to f, they are V\textsubscript{MA}$^-$, V\textsubscript{Pb}$^{2-}$, V\textsubscript{I}$^+$, MA\textsubscript{i}$^+$, Pb\textsubscript{i}$^{2+}$ and I\textsubscript{i}$^-$ respectively. At the centers of the dashed circles, the MA cation, Pb cation and I anion are missing, representing vacancies (a-c). At the centers of the solid circles, an extra MA cation, Pb cation and I anion are incorporated into the MAPbI$_3$ lattice, representing interstitials (d-f).} 
    \label{fig: Optimized structures of different defects}
\end{figure}

The different orientations of MA cations in the MAPbI$_3$ lattice and the tilting and degree of distortion of the PbI$_6$ octahedra in the perovskite lower the symmetry, which in principle gives rise to many nonequivalent positions for defects. Creating a defect by a random choice of site, might not result in the energetically most favorable position. Starting from the tetragonal unit cell of MAPbI$_3$ [Fig. \ref{figure: scheme of creating defects}(a)], there are four MA and Pb cations in the unit cell. Creating one vacancy V\textsubscript{MA} or V\textsubscript{Pb} on each of those four sites gives the same total energy within $\sim 0.05$ eV (Fig. \ref{figure: energy_spread}), indicating that after structure optimization and consequent reorientation of MA cations and PbI$_6$ octahedra, the four sites actually become equivalent. For creating a I vacancy, we consider the six I anions surrounding one Pb cation in an octahedra as inequivalent vacancy sites [Fig. \ref{figure: scheme of creating defects}(a)]. After structure optimization, their total energies give a spread of $\sim 0.3$ eV (Fig. \ref{figure: energy_spread}), so these sites remain inequivalent.

To create interstitials MA\textsubscript{i}, Pb\textsubscript{i} and I\textsubscript{i}, we consider the MAPbI$_3$ cubes [Fig. \ref{figure: scheme of creating defects}(b)]  and insert an interstitial at the center of a cube face, where the six sides of the cube are considered inequivalent. After structure optimization, the interstitals have moved considerably to find their lowest energy position, see the discussion below. Interstitials at different sites present an energy variation of 0.3 to 1.5 eV (Fig. \ref{figure: energy_spread}), indicating the significance of trying out different sites. Fig. \ref{figure: energy_spread} gives the total energies of representative structures of vacancy and interstitial defects. In this work, we use the configurations with the lowest energies for each defect, indicated by red circles, for calculating defect formation energies.

Fig. \ref{fig: Optimized structures of different defects} shows the optimized structures of these lowest energy configurations. The formation of vacancies typically leads to mild changes of the local configurations [Fig. \ref{fig: Optimized structures of different defects} (a-c)]. For instance, a MA vacancy barely affects the surroundings; a Pb vacancy V\textsubscript{Pb}$^{2-}$ leads to a more or less uniform contraction of the six adjacent Pb-I bonds, and an I vacancy V\textsubscript{I}$^+$ pushes the previously bonded Pb cations slightly away. The interstitials generally occupy sites with less symmetry [Fig. \ref{fig: Optimized structures of different defects} (d-f)]. The MA\textsubscript{i}$^+$ occupies the center of a line between two adjacent lattice-site MA cations, and push them slightly aside; the Pb\textsubscript{i}$^{2+}$ adopts a square pyramidal bonding to 5 surrounding I anions; and the I\textsubscript{i}$^-$ typically forms a bridge-like structure together with a lattice-site iodine anion, attracting the two adjacent Pb cations. The MA interstitials show the largest energy variations at different defect sites (Fig. \ref{figure: energy_spread}), because the large size and asymmetrical structure of the MA cation makes a correct positioning in the lattice more critical.

\clearpage

\subsection{Density of defect sites}
In the main text, the term $c_0(D^q)$ in the equation 4 for calculating the defect concentration, defines the density of possible sites for the defect. The value of $c_0(D^q)$ is calculated from 
\begin{equation}
    c_0(D^q) = \frac{n(D^q)}{V_{f.u.}}, 
    \label{eq: density of possible defect sites}
\end{equation}
where the $n(D^q)$ is the number of possible sites for the defect $D^q$ per formula unit of MAPbI$_3$, and the $V_{f.u.}$ is the volume per formula unit of MAPbI$_3$.

The values of $n(D^q)$ for each defect are given in Table \ref{tab: number of possible sites for defect}. For vacancies, counting the number of possible defect sites is fairly straightforward, because these numbers correspond to the atoms in an MAPbI$_3$ cube [Fig. \ref{figure: scheme of creating defects}(b)] that can be removed. For interstitals, counting is based on the configurations of each optimized defective structure shown in Fig. \ref{fig: Optimized structures of different defects} (d-f).

\begin{table}[h!]
    \caption{Number of possible sites for each type of defect and the  counting rule used.}
    \label{tab: number of possible sites for defect}
    \centering
    \begin{ruledtabular}    
    \begin{tabular}{p{0.08\linewidth} p{0.08\linewidth} p{0.8\linewidth}}
        Defect & $n(D^q)$ & Remarks \\
        \hline
        V\textsubscript{MA}$^-$ & 1 & 1 MA cation in the center of the MAPbI$_3$ cube [Fig. \ref{figure: scheme of creating defects}(b)].\\
        V\textsubscript{Pb}$^{2-}$ & 1 & 8 Pb cations at the corners of the cube [Fig. \ref{figure: scheme of creating defects}(b)], with each corner shared by 8 cubes; so 8/8 = 1.\\
        V\textsubscript{I}$^+$ & 3 & 12 iodine anions in the middle of each edge [Fig. \ref{figure: scheme of creating defects}(b)], with each edge shared by 4 cubes; so 12/4=3. \\
        MA\textsubscript{i}$^+$ & 3 & the MA interstitial occupies a face of a MAPbI$_3$ cube [Fig. \ref{fig: Optimized structures of different defects}(d)]; with 6 faces per cube, and each face shared by 2 cubes, this means 6/2 = 3. \\
        Pb\textsubscript{i}$^{2+}$ & 12 & the Pb interstitial has a square pyramidal bonding to 5 surrounding I anions [Fig. \ref{fig: Optimized structures of different defects}(e)]; all 12 iodine anions in the middle of the edges can serve as an apex of a pyramid, and all pyramids are inside the cube.\\
        I\textsubscript{i}$^-$ & 3 & the iodine interstitial occupies a site next to a lattice iodine anion in the middle of each edge [Fig. \ref{fig: Optimized structures of different defects}(f)]; with each edge shared by 4 cubes, one has 12/4 = 3.\\
    \end{tabular}
    \end{ruledtabular}
\end{table}

\clearpage

\section{DFEs and CSTLs from PBE-optimized structures}

\begin{figure} [h]
    \includegraphics[width=0.6\textwidth]{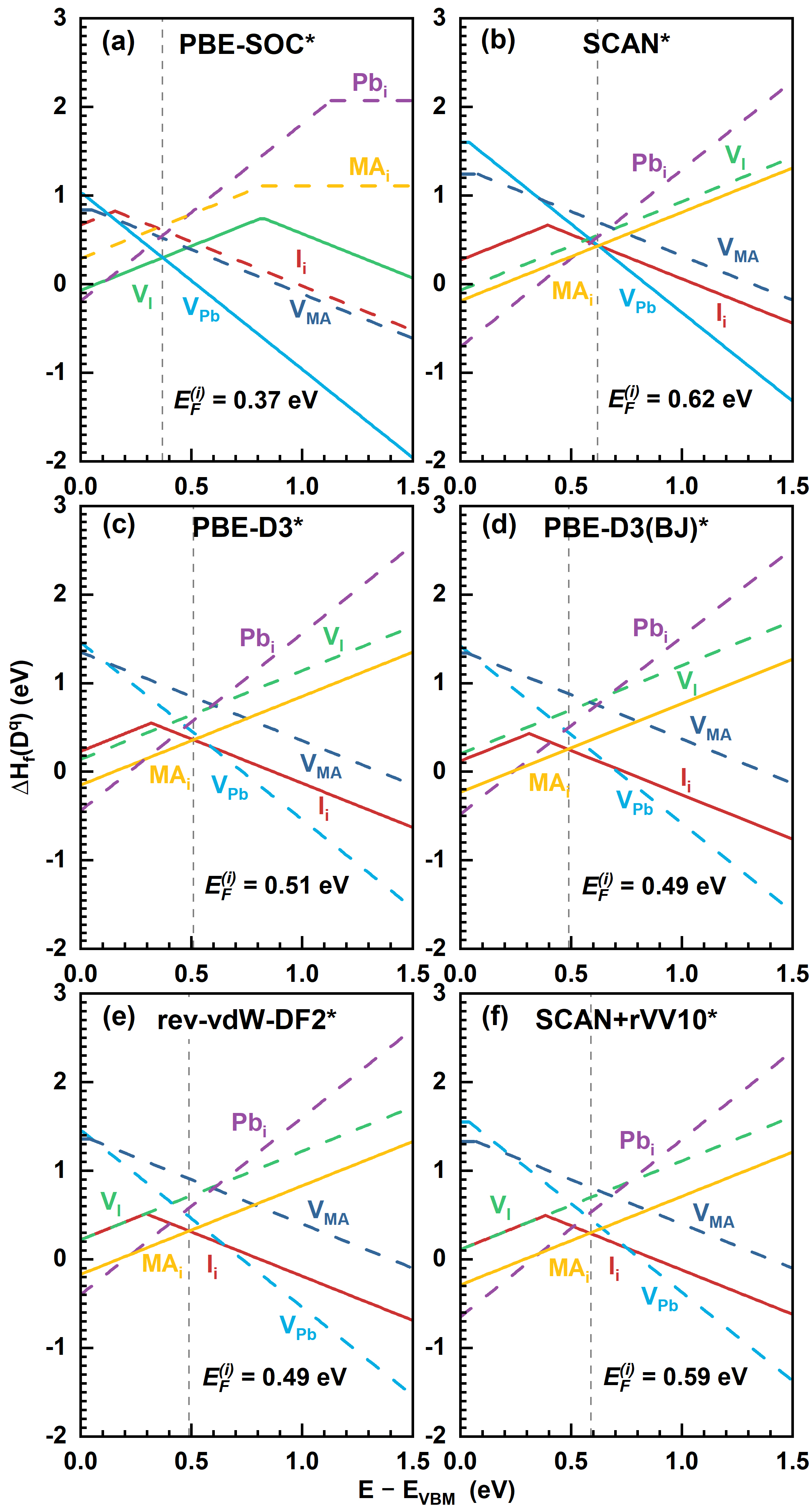}
    \caption{Defect formation energies (DFEs) calculated by using different functionals, but based upon structures optimized using PBE.  \label{figure:DFE_PBE_combine}}
\end{figure}

\begin{figure} [h]
    \includegraphics[width=0.6\textwidth]{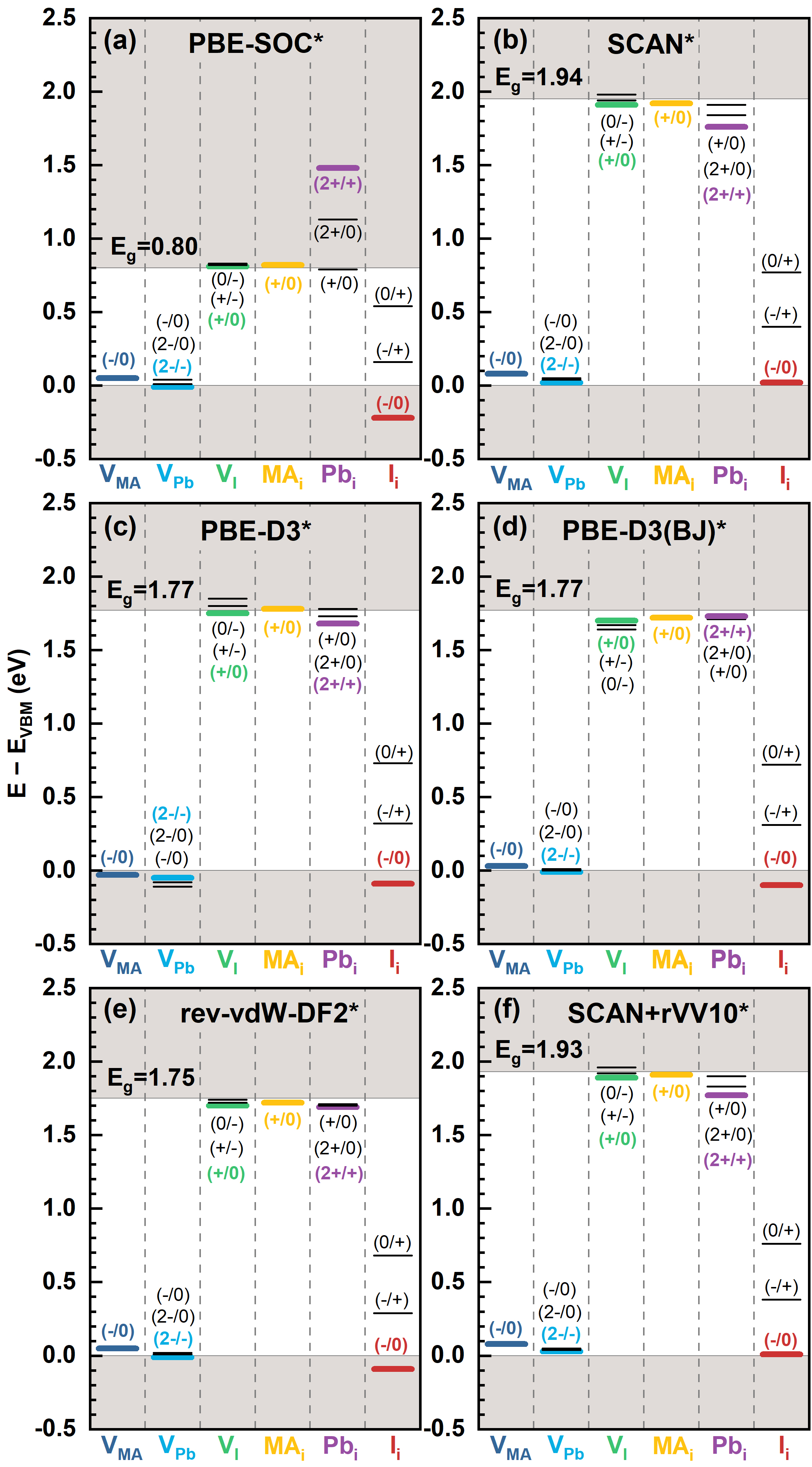}
    \caption{Charge state transition levels (CSTLs) calculated by using different functionals, but based upon structures optimized using PBE. \label{figure:TDTL_PBE_combine}}
\end{figure}

\clearpage

\section{Chemical Potentials}

\begin{table}[h]
    \caption{Chemical potentials of the Pb and I at different growth conditions calculated by using different functionals. The chemical potentials, $\Delta \mu$, of I and Pb are relative values, which are referred to the I$_2$ molecule and Pb bulk, respectively. The asterisk * on top of PBE-SOC represent the single point calculations using PBE-SOC are based on the equilibrium structures optimized using PBE.}
    \label{table: Chemical potentials}
    \begin{ruledtabular}
    \begin{tabular}{lcccccc}
        \multirow{2}{*}{Functional} & \multicolumn{2}{c}{Iodine-rich} & \multicolumn{2}{c}{Iodine-medium} & \multicolumn{2}{c}{Iodine-poor}\\
        \cline{2-3} \cline{4-5} \cline{6-7}
          & $\Delta \mu _I$ (eV) & $\Delta \mu _{Pb}$ (eV) & $\Delta \mu _I$ (eV) & $\Delta \mu _{Pb}$ (eV) & $\Delta \mu _I$ (eV) & $\Delta \mu _{Pb}$ (eV)\\
        \hline
        PBE & 0.00 & $-2.39$ & $-0.60$ & $-1.20$ & $-1.20$ & 0.00\\
        PBE-SOC* & 0.00 & $-2.20$ & $-0.55$ & $-1.10$ & $-1.10$ & 0.00\\
        SCAN & 0.00 & $-2.50$ & $-0.63$ & $-1.25$ & $-1.25$ & 0.00\\
        PBE-D3 & 0.00 & $-2.78$ & $-0.69$ & $-1.39$ & $-1.39$ & 0.00\\
        PBE-D3(BJ) & 0.00 & $-2.81$ & $-0.70$ & $-1.40$ & $-1.40$ & 0.00\\
        rev-vdW-DF2 & 0.00 & $-2.80$ & $-0.70$ & $-1.40$ & $-1.40$ & 0.00\\
        SCAN+rVV10 & 0.00 & $-2.68$ & $-0.67$ & $-1.34$ & $-1.34$ & 0.00\\
    \end{tabular}
    \end{ruledtabular}
\end{table}

\clearpage

\section{Neutral Defects}
\begin{table}[h]
    \caption{Formation energies $\Delta H_f$ (eV) and concentrations $c$ (cm$^{-3}$) of neutral defects in MAPbI$_3$ calculated using different functionals.}
    \label{table: DFE_neutral }
    \begin{ruledtabular}
    \begin{tabular}{lllllll}
        Functional &V\textsubscript{MA}$^0$ & V\textsubscript{Pb}$^{0}$ & V\textsubscript{I}$^0$ & MA\textsubscript{i}$^0$ & Pb\textsubscript{i}$^{0}$ & I\textsubscript{i}$^0$\\
        \hline
\multicolumn{7}{l}{Defect formation energy $\Delta H_f$ (eV) \footnotemark[1]}\\
    PBE & 1.17 & 1.70 & 1.53 & 1.90 & 2.97 & 1.29\\
    PBE-SOC* & 0.84 & 1.02 & 0.74 & 1.11 & 2.07 & 1.20\\
    SCAN & 1.07 & 1.33 & 1.82 & 1.94 & 3.20 & 1.16\\
    PBE-D3 & 1.14 & 1.15 & 1.80 & 1.76 & 3.13 & 0.85\\
    PBE-D3(BJ) & 1.18 & 1.14 & 1.87 & 1.71 & 3.16 & 0.77\\
    rev-vdW-DF2 & 1.22 & 1.17 & 1.85 & 1.68 & 3.15 & 0.80\\
    SCAN+rVV10 & 1.09 & 1.15 & 1.95 & 1.88 & 3.25 & 1.00\\
\multicolumn{7}{l}{Defect concentration $c$ (cm$^{-3}$) \footnotemark[1]}\\
PBE & 7.95$\times 10^{1}$ & 1.15$\times 10^{-7}$ & 2.31$\times 10^{-4}$ & 1.58$\times 10^{-10}$ & 1.23$\times 10^{-28}$ & 2.51$\times 10^{0}$\\
PBE-SOC* & 2.62$\times 10^{7}$ & 3.33$\times 10^{4}$ & 4.26$\times 10^{9}$ & 2.14$\times 10^{3}$ & 1.62$\times 10^{-13}$ & 6.64$\times 10^{1}$\\
SCAN & 4.83$\times 10^{3}$ & 2.01$\times 10^{-1}$ & 2.66$\times 10^{-9}$ & 2.65$\times 10^{-11}$ & 2.40$\times 10^{-32}$ & 4.22$\times 10^{2}$\\
PBE-D3 & 2.95$\times 10^{2}$ & 2.24$\times 10^{2}$ & 6.93$\times 10^{-9}$ & 3.44$\times 10^{-8}$ & 3.69$\times 10^{-31}$ & 6.99$\times 10^{7}$\\
PBE-D3(BJ) & 7.07$\times 10^{1}$ & 3.36$\times 10^{2}$ & 4.63$\times 10^{-10}$ & 2.39$\times 10^{-7}$ & 8.69$\times 10^{-32}$ & 1.50$\times 10^{9}$\\
rev-vdW-DF2 & 1.48$\times 10^{1}$ & 7.70$\times 10^{1}$ & 8.29$\times 10^{-10}$ & 7.08$\times 10^{-7}$ & 1.42$\times 10^{-31}$ & 5.00$\times 10^{8}$\\
SCAN+rVV10 & 2.11$\times 10^{3}$ & 1.90$\times 10^{2}$ & 2.45$\times 10^{-11}$ & 3.26$\times 10^{-10}$ & 1.22$\times 10^{-32}$ & 1.80$\times 10^{5}$\\
    \end{tabular}
    \end{ruledtabular}
    \footnotetext[1]{Here, the defect formation energies are rather large while the defect concentrations are fairly low because these are for neutral defects. The defects are usually stable in their corresponding charged states, which are discussed in the main text.}
\end{table}

\clearpage

\bibliography{Supplemental_material}

\bibliographystyle{apsrev4-2}